\documentclass[fleqn, 10pt]{wlscirep}
\usepackage[utf8]{inputenc}
\usepackage[T1]{fontenc}
\usepackage{colortbl}
\usepackage{bm}
\usepackage{marvosym}

\title{A Versatile Pathology Co-pilot via Reasoning Enhanced Multimodal Large Language Model}
\author[1]{Zhe Xu}
\author[1]{Ziyi Liu}
\author[1]{Junlin Hou}
\author[1]{Jiabo Ma}
\author[1]{Cheng Jin}
\author[1]{Yihui Wang}
\author[1]{Zhixuan Chen}
\author[3]{Zhengyu Zhang} 
\author[1]{Fuxiang Huang}
\author[1]{Zhengrui Guo}
\author[1]{Fengtao Zhou}
\author[1]{Yingxue Xu}
\author[1]{Xi Wang}
\author[2]{Ronald Cheong Kin Chan}
\author[3,4,5]{Li Liang}
\author[1,6,7,8,9 \Letter]{Hao Chen}

\affil[1]{Department of Computer Science and Engineering, Hong Kong University of Science and Technology, Hong Kong, China}

\affil[2]{Department of Anatomical and Cellular Pathology, Chinese University of Hong Kong, Hong Kong, China}
\affil[3]{Department of Pathology, Nanfang Hospital and School of Basic Medical Sciences, Southern Medical University, Guangzhou, China.}
\affil[4]{Guangdong Provincial Key Laboratory of Molecular Tumor Pathology, Guangzhou, China.}
\affil[5]{Jinfeng Laboratory, Chongqing, China}
\affil[6]{Department of Chemical and Biological Engineering, Hong Kong University of Science and Technology, Hong Kong SAR, China}
\affil[7]{Division of Life Science, Hong Kong University of Science and Technology, Hong Kong SAR, China}
\affil[8]{State Key Laboratory of Nervous System Disorders, The Hong Kong University of Science and Technology, Hong Kong SAR, China}
\affil[9]{HKUST Shenzhen-Hong Kong Collaborative Innovation Research Institute, The Hong Kong University of Science and Technology, Futian, Shenzhen, China}

\affil[]{\textbf{Corresponding author: Hao Chen  (jhc@cse.ust.hk)}}

\begin{abstract}
Multimodal large language models (MLLMs) have emerged as powerful tools for computational pathology, offering unprecedented opportunities to integrate pathological images with language context for comprehensive diagnostic analysis. These models hold particular promise for automating complex tasks that traditionally require expert interpretation of pathologists. 
However, current MLLM approaches in pathology demonstrate significantly constrained reasoning capabilities, primarily due to their reliance on expensive chain-of-thought annotations. Additionally, existing methods remain limited to simplex application of visual question answering (VQA) at the region-of-interest (ROI) level, failing to address the full spectrum of diagnostic needs such as ROI classification, detection, segmentation, whole-slide-image (WSI) classification and VQA in clinical practice.
In this study, we present SmartPath-R1, a versatile MLLM capable of simultaneously addressing both ROI-level and WSI-level tasks while demonstrating robust pathological reasoning capability. Our framework combines scale-dependent supervised fine-tuning and task-aware reinforcement fine-tuning, which circumvents the requirement for chain-of-thought supervision by leveraging the intrinsic knowledge within MLLM. Furthermore, SmartPath-R1 integrates multiscale and multitask analysis through a mixture-of-experts mechanism, enabling dynamic processing for diverse tasks. We curate a large-scale dataset comprising 2.3M ROI samples and 188K WSI samples for training and evaluation. Extensive experiments across 72 tasks validate the effectiveness and superiority of the proposed approach. This work represents a significant step toward developing versatile, reasoning-enhanced AI systems for precision pathology.
\end{abstract}

\begin{document}

    \maketitle

\section{Introduction}
Pathology, as the gold standard in modern medicine \cite{he2024foundation,ikezogwo2023quilt,hou2024self,hou2025qmix}, has undergone a digital revolution that allows computational approaches to improve diagnostic precision. Although early computational pathology models \cite{chen2016dcan,uni} show promise in unimodal image analysis, contemporary diagnostic practice demands multimodal integration of visual patterns with the language context \cite{huang2023visual,huang2022adversarial,mormont2024visual,xu2022hisa,xu2022point,conch,mstar,xu2023bilateral,xu2024exploiting,huang2024gradient,huang2024dynamic,xiang2025vision,chen2025segment}.

\begin{figure*}
    \centering
    \includegraphics[width=1\linewidth]{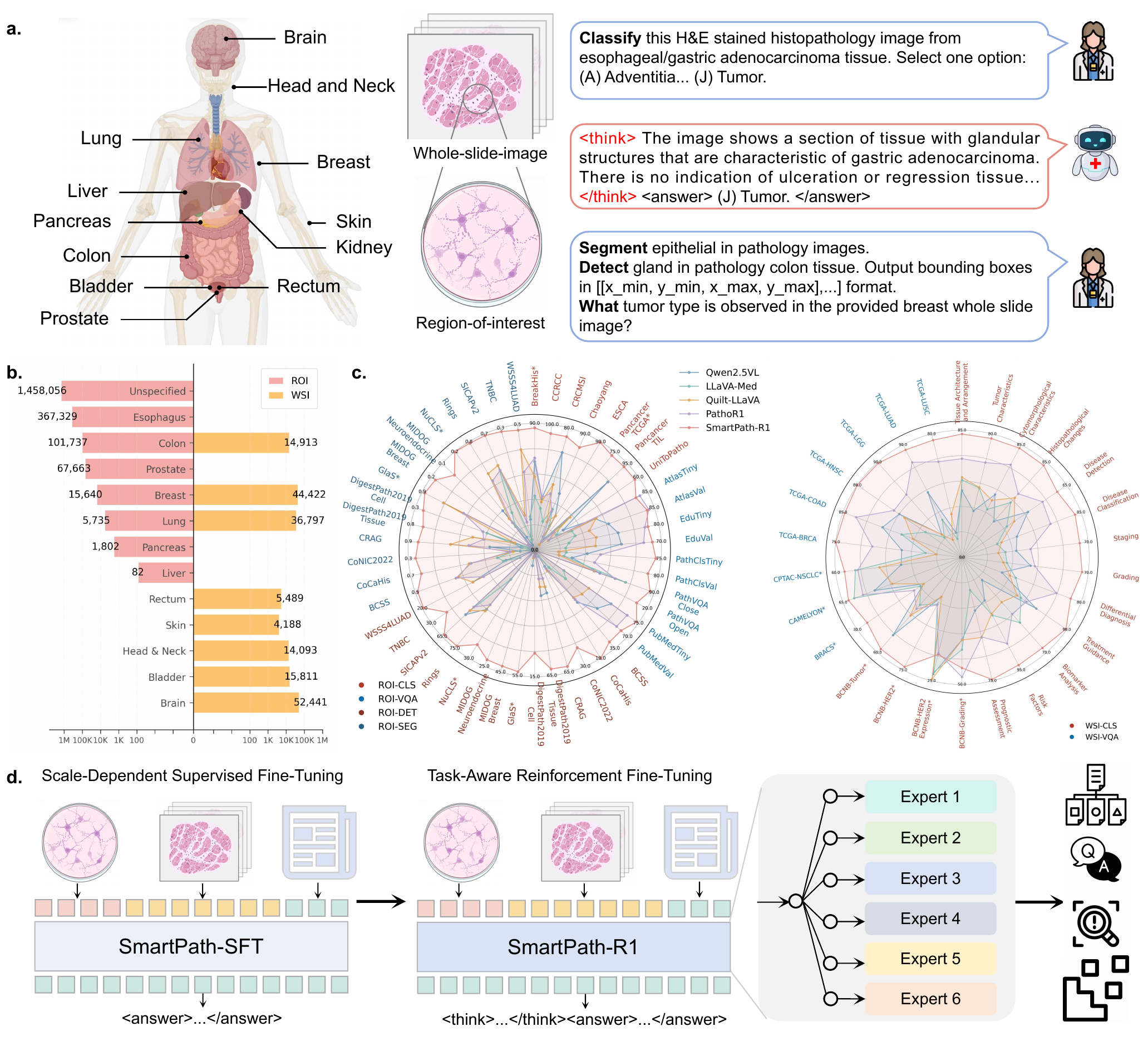}
\caption{\textbf{Overview of the SmartPath-R1.} \textbf{a.} SmartPath-R1 is a versatile pathology co-pilot that can simultaneously address both ROI-level and WSI-level tasks across various anatomical regions while demonstrating robust pathological reasoning capability. \textbf{b.} We curate a dataset comprising 2,295,095 ROI-level and 188,154 WSI-level samples, enabling comprehensive model training and evaluation. \textbf{c.} Performance evaluation of MLLMs across 46 ROI-level tasks and 26 WSI-level tasks.  * represents external validation tasks. SmartPath-R1 demonstrates exceptional performance over state-of-the-art MLLMs. \textbf{d.} The framework, comprising scale-dependent supervised fine-tuning and task-aware reinforcement fine-tuning, enables multimodal pathological reasoning within <think></think> tags. ROIs and WSIs are encoded into variable-length visual token sequences. Mixture-of-expert mechanism is used to address a spectrum of tasks, from ROI-level classification to WSI-level VQA.}
\label{fig:overall}
\end{figure*}

The emergence of multimodal large language models (MLLMs)~\cite{liu2024improved,liu2023visual,pathchat,quiltllava} has created new opportunities, yet two fundamental limitations persist.
First, existing models often fail to achieve clinically meaningful reasoning. This is largely due to their reliance on supervised fine-tuning ~\cite{pathasst,pathchat,quiltllava}, which optimizes the accuracy of prediction but fails to capture the sequential decision-making process inherent in pathological diagnosis, where experts progressively integrate visual cues and knowledge. As a result, models may learn superficial correlations rather than the critical reasoning pathways needed for complex cases.
Second, existing approaches exhibit limited versatility in handling various computational pathology tasks. Although current MLLMs demonstrate preliminary capabilities in ROI-level VQA~\cite{pathchat,quiltllava}, their capacity to concurrently process multiscale queries (ROI and WSI level) remains underdeveloped. These methods lack unified frameworks to integrate fundamental pathological tasks, including fine-grained classification, detection, and segmentation, with diagnostic reasoning. This fragmentation hinders the development of comprehensive diagnostic assistants capable of mirroring the multitask analytical workflows of pathologists, where visual interpretation and semantic reasoning are intrinsically coupled across spatial hierarchies.

To overcome these limitations, we present SmartPath-R1 (Figure \ref{fig:overall}), an MLLM trained on extensive datasets comprising 1,964,121 ROI samples and 179,897 WSI samples, with a rigorous evaluation performed on 330,974 ROI and 8,257 WSI test samples. Our approach uniquely embeds diagnostic reasoning by formulating pathology interpretation as a sequential policy optimization process, where the model learns optimal evidence-gathering policies through reward signals mimicking pathologist decision trajectories. Crucially, our approach eliminates the need for expensive chain-of-thought supervision by mining the knowledge inherent in MLLMs, thereby enhancing scalability. Moreover, SmartPath-R1 unifies multiscale and multitask analysis via a mixture-of-expert mechanism that dynamically handles ROI-level morphometrics to WSI-level topological features for different tasks.

This paradigm shift moves beyond current supervised learning constraints, demonstrating for the first time how AI systems can acquire human-like diagnostic reasoning without procedural annotation and significantly outperform state-of-the-art multimodal approaches on various pathological tasks.

\begin{figure*}
    \centering
    \includegraphics[width=1\linewidth]{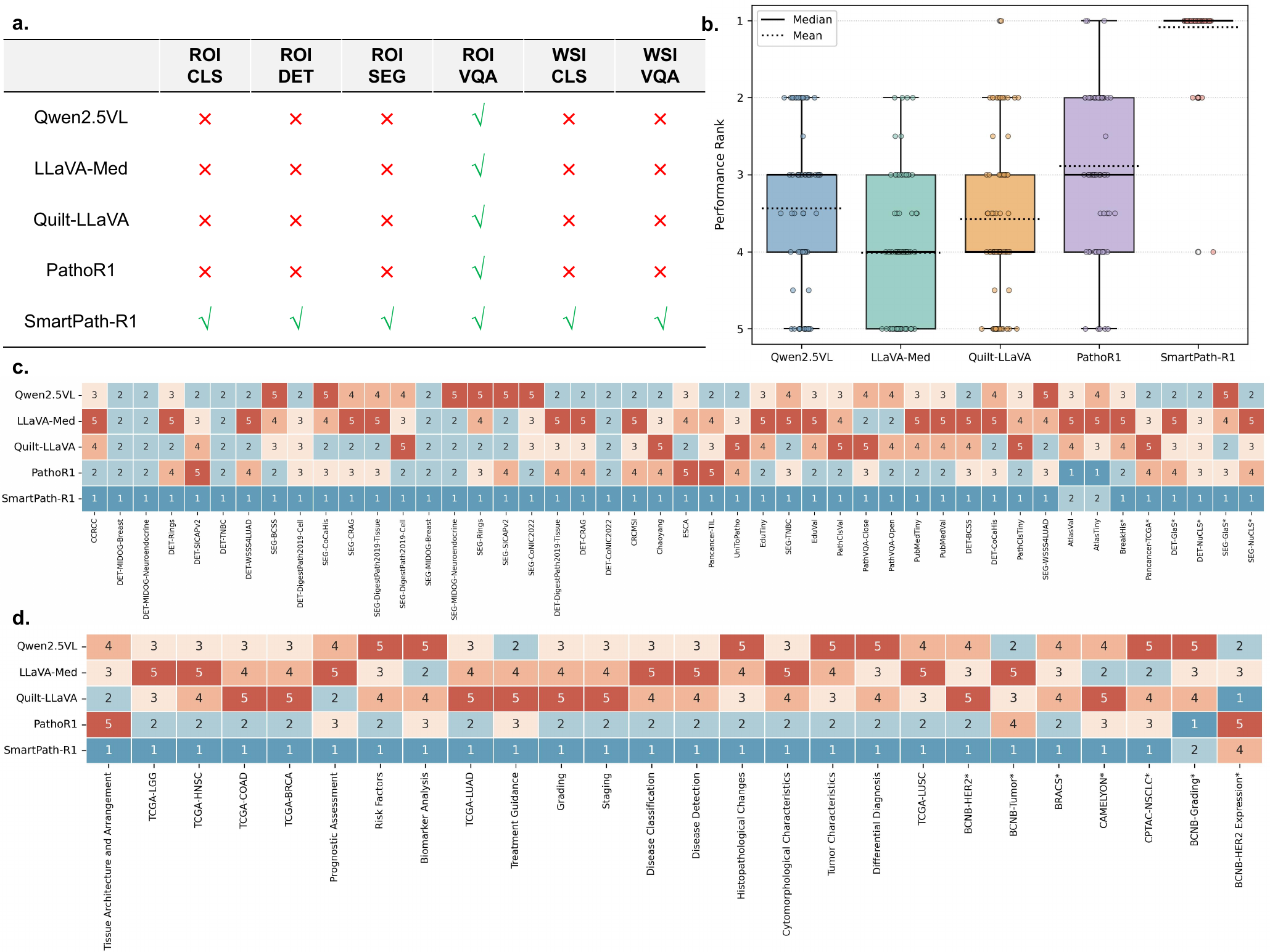}
\caption{\textbf{Comprehensive Comparison of MLLMs across 72 Tasks.} \textbf{a.} Task types evaluated by different MLLMs. CLS, DET, SEG, and VQA represent classification, detection, segmentation, and visual question answering, respectively. \textbf{b.} Average rank of MLLMs across 72 downstream tasks. The solid line and the dashed line represent the median and the mean, respectively. \textbf{c.} Ranking order of MLLMs across 46 ROI-level tasks. \textbf{d.} Ranking order of MLLMs on 26 WSI-level tasks. * represents external validation datasets. If a model achieves the best performance, its rank value is set to 1. Models with identical metric values receive the same rank (tied ranking).}
\label{fig:overall1}
\end{figure*}

\section{Results}

We evaluate various MLLMs on 72 tasks, encompassing 8 ROI-level classification tasks, 14 ROI-level pathological detection tasks, 14 ROI-level pathological segmentation tasks, 10 ROI-level VQA tasks, 9 WSI-level classification tasks, and 17 WSI-level VQA tasks.
Since the tasks involved different types of evaluation metrics, we evaluate the overall performance of the MLLMs using an average ranking approach. The model with the best performance was ranked 1st, while the model with the lowest performance was ranked 5th. As shown in Figure \ref{fig:overall1}, the SmartPath-R1 model achieves the top average rank score of 1.1 (ranked first in 68 tasks), outperforming the second-best model, PathoR1, which has a ranking score of 2.9 (ranked first in 3 tasks). The results clearly indicate that SmartPath-R1 achieves state-of-the-art performance and is much more generalizable compared to the other MLLMs. 

\subsection*{ROI-Level Classification}

Pathological ROI-level classification is a fundamental task in computational pathology, aiming to automatically categorize pathology ROI images into predefined classes. Unlike generic image classification, this task demands strong reasoning capabilities to decipher subtle morphological patterns (e.g., nuclear pleomorphism, tissue architecture) and integrate domain-specific knowledge (e.g., grading criteria, staining artifacts).  
Although traditional deep learning approaches excel at pattern recognition, integrating explicit reasoning mechanisms is increasingly vital to bridge the gap between AI and clinical trustworthiness.

\begin{figure*}
    \centering
    \includegraphics[width=1\linewidth]{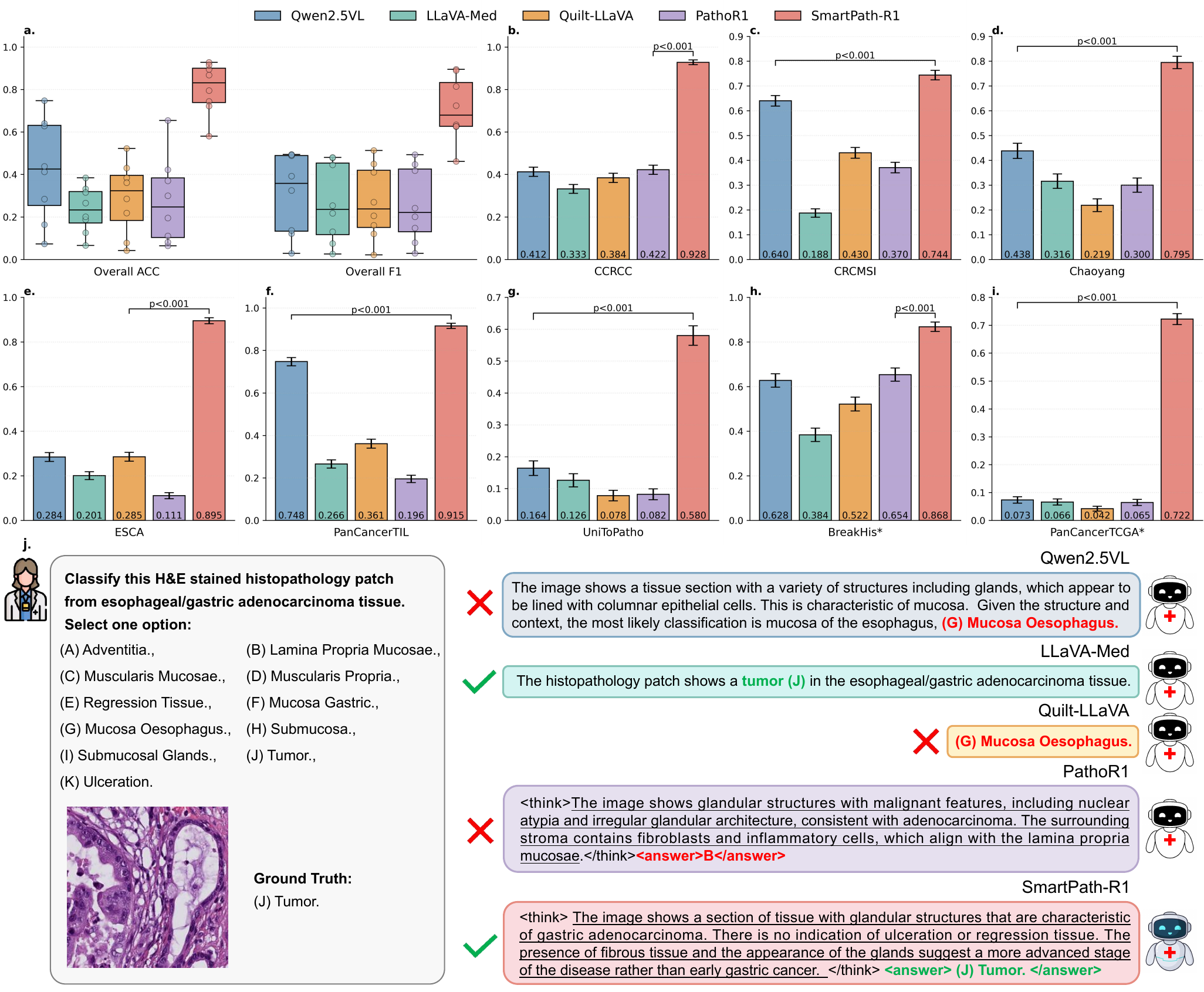}
\caption{\textbf{Performance of MLLMs on ROI-Level Classification Tasks.}
\textbf{a.} Average performance of MLLMs based on accuracy (ACC) and F1 across 8 ROI-level classification tasks.
\textbf{b-i.} Model performance on specific tasks. * represents external validation datasets. Error bars represent 95\% CI. The box limits represent the standard error. P-values are computed using a Wilcoxon signed-rank two-sided test \cite{demvsar2006statistical}. Additional results are shown in Extended Data Table  \ref{tab:ROI:avg_cls} and Tables \ref{tab:ROI:CCRCC}-\ref{tab:ROI:PanCancerTCGA}. \textbf{j.} An example of ROI-level classification along with the results generated by various MLLMs.
}
\label{fig:ROI_cls}
\end{figure*}

In our comprehensive evaluation, SmartPath-R1 is benchmarked against four state-of-the-art MLLMs: Qwen2.5VL \cite{bai2025qwen2} (general purpose vision-language model), LLaVA-Med \cite{li2023llava} (biomedical-focused MLLM), Quilt-LLaVA \cite{quiltllava} and PathoR1 \cite{zhang2025patho} (specialized for pathology).
As shown in Figure \ref{fig:ROI_cls}a-i, the evaluation of MLLMs across 8 distinct tasks, including 2 external validation datasets, reveals important insights into their capabilities.  
Qwen2.5VL surprisingly outperforms several domain-specialized models across multiple pathological datasets. This finding can be attributed to its extensive pretraining on diverse vision-language tasks, which fosters more robust feature representations capable of generalizing to unseen pathological patterns. PathoR1's relatively limited performance highlights the challenges of overspecialization in pathological AI systems. While designed specifically for ROI-level VQA, its constrained training scope and modest sample size render it less adaptable to broader diagnostic tasks.
In contrast, SmartPath-R1 achieves particularly strong results, where it outperforms the second-best approach by 38.3\% in average accuracy. This superior performance can be attributed to the models' ability to capture tissue patterns and contextual relationships in these complex classification tasks.

We present an example of classification task in Figure \ref{fig:ROI_cls}j.  General purpose Qwen2.5VL struggles with domain-specific reasoning, as it prioritizes broad visual-language alignment over histopathologic logic. LLaVA-Med's correct but terse output (``shows a tumor'') lacks explanatory depth, limiting its utility in clinical settings where justification is paramount. Quilt-LLaVA and PathoR1's misclassification as (G) and (B) likely stems from isolated feature recognition without integrating stromal or nuclear atypia cues. SmartPath-R1 identifies the tissue as (J) Tumor, reasoning that the glandular morphology indicates advanced gastric adenocarcinoma, excluding early-stage or benign features (e.g., ulceration or regression tissue). This aligns with the clinical context of adenocarcinoma, where malignant glandular presence is a hallmark.

\subsection*{ROI-Level Detection}
\begin{figure*}
    \centering
    \includegraphics[width=1\linewidth]{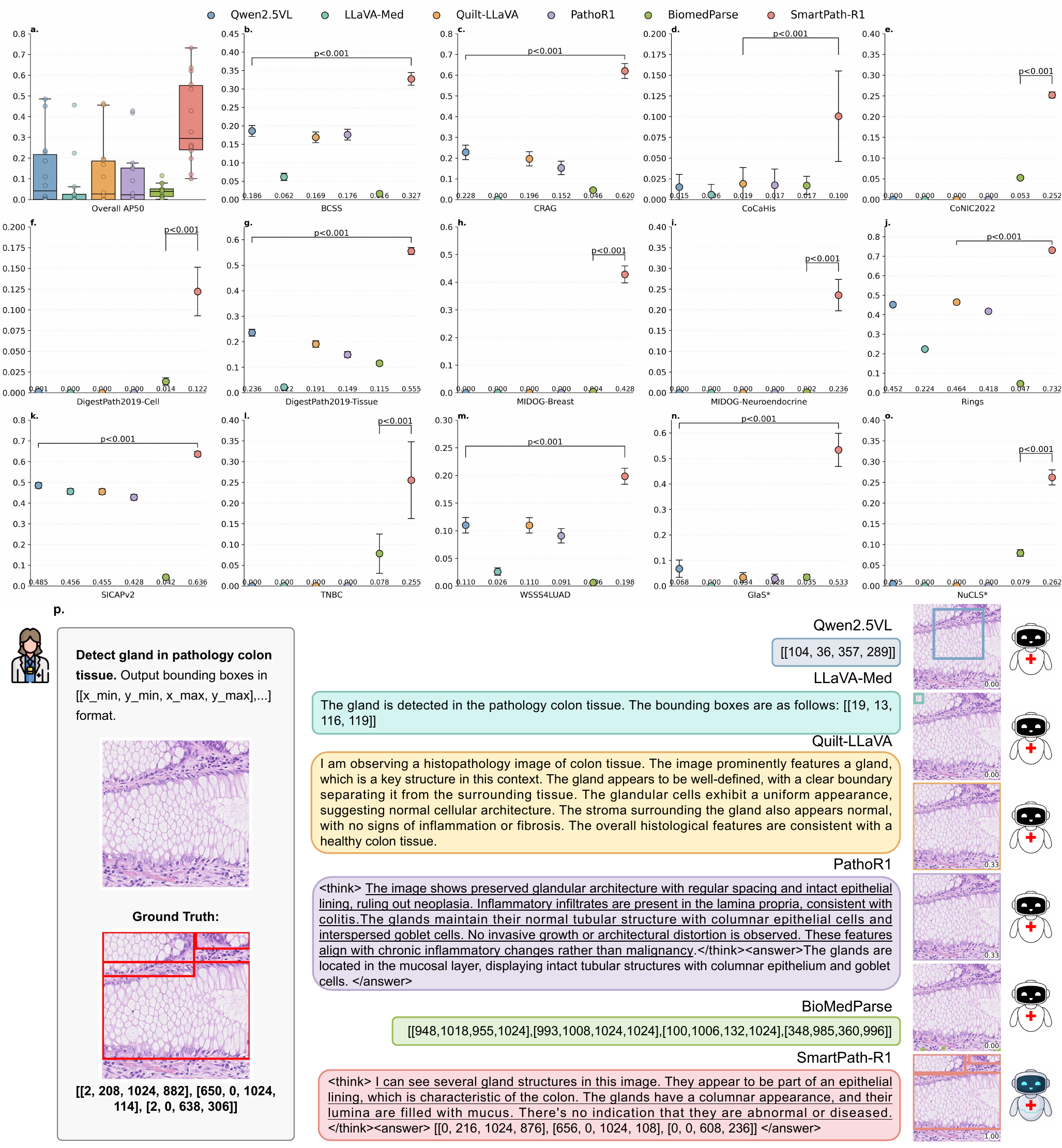}
\caption{\textbf{Performance of MLLMs on ROI-Level Detection Tasks.}
\textbf{a.} Average performance of MLLMs based on average precision at IoU threshold 0.5 (AP50) across 14 ROI-level detection tasks.
\textbf{b-o.} Model performance on specific tasks. * represents external validation datasets. Error bars represent 95\% CI. The box limits represent the standard error. P-values are computed using a Wilcoxon signed-rank two-sided test. Additional results are shown in Extended Data Table \ref{tab:ROI_det:avg}, Tables \ref{tab:ROI_det:BCSS}-\ref{tab:ROI_det:NuCLS}, Figure \ref{fig:ROI_det_30} and Figure \ref{fig:ROI_det_70}. \textbf{p.} An example of ROI-level detection task along with the results generated by various MLLMs. Red rectangles indicate ground truth annotations, while predicted bounding boxes of each method are shown in their corresponding colors. The AP50 scores for each method are displayed in the lower-right corner, quantitatively demonstrating their detection accuracy relative to the ground truth.
}
\label{fig:ROI_det}
\end{figure*}

ROI-level detection involves the localization and classification of pathological entities (e.g., cells, nuclei, tumor regions) in images. This task serves as the foundation for automated diagnosis, prognostic assessment, and biomedical research, enabling high-throughput analysis of morphological features that are often imperceptible to human observers. 
The clinical significance of this task necessitates not only detection accuracy but also interpretive reasoning—the ability to contextualize morphological patterns within pathological paradigms. For instance, distinguishing malignant nuclei from benign counterparts requires reasoning about spatial relationships (e.g., tumor infiltration), texture heterogeneity, and diagnostic criteria. Traditional deep learning approaches, while proficient in pattern recognition, often lack explicit reasoning mechanisms.

Figure \ref{fig:ROI_det} presents a comparative evaluation of SmartPath-R1 with existing MLLMs and a state-of-the-art medical image detection model BiomedParse. The analysis leverages 14 benchmark datasets spanning nuclear, cell, and tissue detection, focusing on Average Precision at an IoU threshold of 0.5 (AP50) as the primary metric. The results demonstrate that SmartPath-R1 achieves a significantly higher average AP50 across all datasets (Figure \ref{fig:ROI_det}a), indicating superior detection performance. Dataset-specific performances (Figures \ref{fig:ROI_det}b-o) further reveal that SmartPath-R1 exhibits particularly strong performance on all the datasets (p < 0.001).  This suggests that SmartPath-R1 possesses a greater capacity for reasoning about the complex visual cues necessary for accurate detection.

The qualitative examples are shown in Figure \ref{fig:ROI_det}p. SmartPath-R1's bounding boxes tightly enclose the glands, demonstrating a precise understanding of gland boundaries and morphology. In contrast, other models often produce inaccurate or incomplete bounding boxes, indicating a failure to reason about the subtle visual cues that distinguish glands from surrounding tissue. These findings suggest that SmartPath-R1's enhanced reasoning capabilities contribute to its superior performance on ROI-level detection tasks, enabling it to accurately localize objects even in challenging histological contexts.

\subsection*{ROI-Level Segmentation}

\begin{figure*}
    \centering
    \includegraphics[width=1\linewidth]{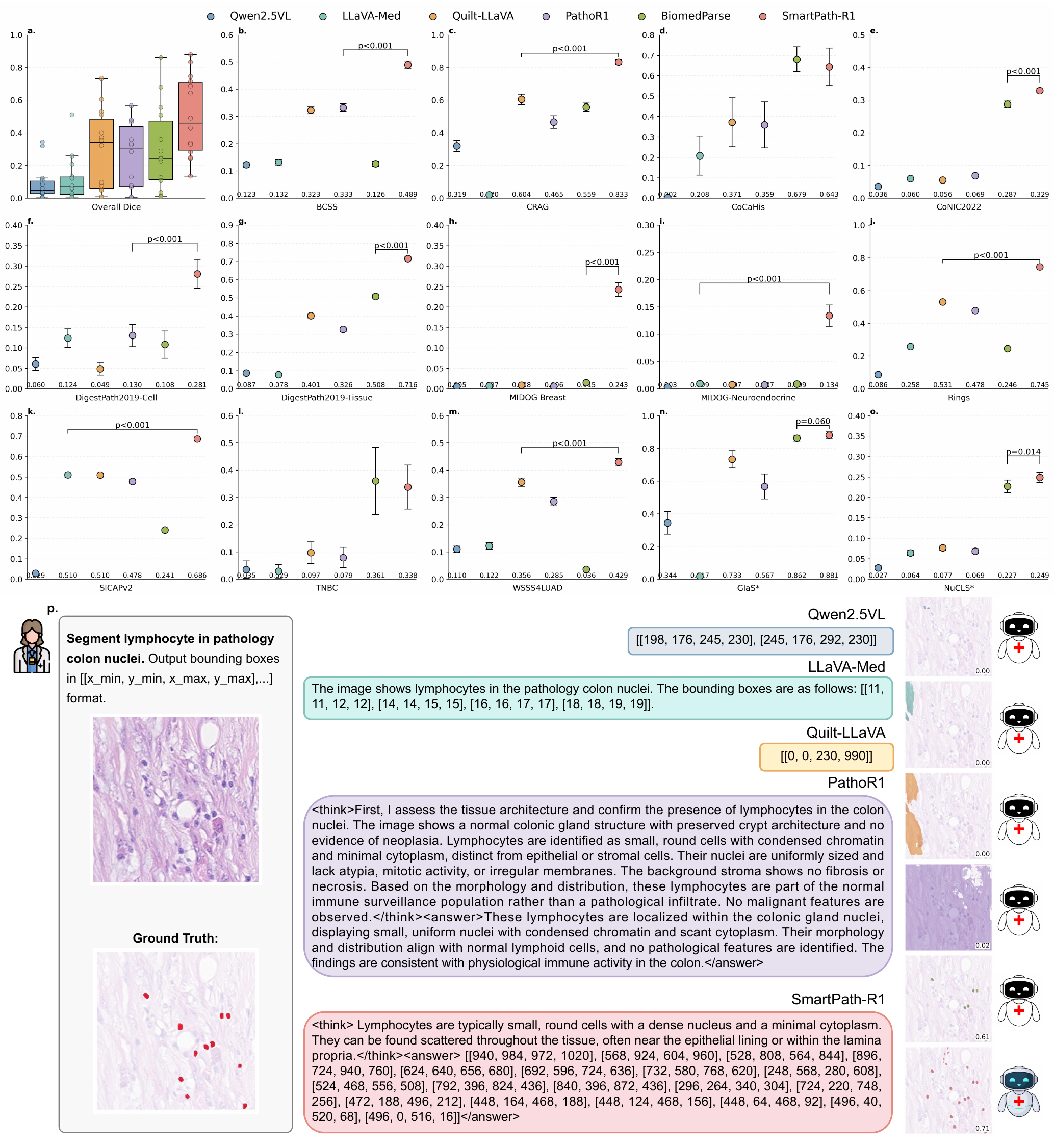}
\caption{\textbf{Performance of MLLMs on ROI-Level Segmentation Tasks.} For MLLMS, masks are generated by using the predicted objects' coordinates as prompts for the pre-trained MedSAM model.
\textbf{a.} Average performance of MLLMs based on Dice score across 14 ROI-level segmentation tasks.
\textbf{b-o.} Model performance on specific tasks. * represents external validation datasets. P-values are computed using a Wilcoxon signed-rank two-sided test. Additional results are shown in Extended Data Table \ref{tab:ROI_seg:avg} and Tables \ref{tab:ROI_seg:BCSS}-\ref{tab:ROI_seg:GlaS}. \textbf{p.} An example of ROI-level segmentation task, along with the results generated by various MLLMs. Red masks indicate ground truth annotations, while predicted masks of each method are shown in their corresponding colors. The Dice scores for each method are displayed in the lower-right corner, quantitatively demonstrating their segmentation accuracy relative to the ground truth.
}
\label{fig:ROI_seg}
\end{figure*}

Pathological ROI-level segmentation is a critical task in computational pathology that involves pixel-wise classification of histological structures in images. This fine-grained delineation enables quantitative analysis of tissue morphology, which is indispensable for objective diagnosis, cancer grading, and biomarker assessment. 
Unlike natural image segmentation, pathological segmentation presents unique challenges due to staining heterogeneity, overlapping cellular structures, and ambiguous pathological boundaries, demanding algorithms that combine low-level feature extraction with high-level diagnostic reasoning.
While deep learning has advanced segmentation accuracy, clinical deployment requires systems capable of pathology-informed reasoning to interpret morphological patterns in diagnostically meaningful contexts.

To assess the segmentation capabilities of SmartPath-R1, we conduct a comparative evaluation against existing models on 14 ROI-level segmentation datasets (Figure \ref{fig:ROI_seg}). For segmentation using MLLMs, the predicted objects' coordinates are used as prompts for the pre-trained MedSAM \cite{ma2024segment} model to generate masks. The primary evaluation metric is the Dice score, which quantifies the overlap between predicted and ground truth segmentation masks.
The results demonstrate that SmartPath-R1 achieves a significantly higher average Dice score across all datasets (Figures \ref{fig:ROI_seg}a) compared to other models, indicating superior segmentation performance. Analysis of dataset-specific performance (Figures \ref{fig:ROI_seg}b-o) reveals that SmartPath-R1 exhibits particularly strong performance on datasets such as BCSS, CoNIC2022, CRAG, DigestPath2019 Cell, DigestPath2019 Tissue, MIDOG Breast, MIDOG Neuroendocrine, Rings, SICAPv2, and WSS4LUAD (p < 0.001). This suggests that SmartPath-R1 possesses a greater capacity for reasoning about the complex visual cues necessary for accurate segmentation.

Figure \ref{fig:ROI_seg}p presents the qualitative examples. SmartPath-R1's segmentation masks closely align with the lymphocyte boundaries, demonstrating a precise understanding of cell morphology. In contrast, other models often produce inaccurate or incomplete masks, indicating a failure to reason about the subtle visual cues that distinguish lymphocytes from surrounding tissue. These findings suggest that SmartPath-R1's enhanced reasoning capabilities contribute to its superior performance on ROI-level segmentation tasks, enabling it to accurately delineate histological structures even in complex pathology images.

\subsection*{ROI-Level VQA}
\begin{figure*}
    \centering
    \includegraphics[width=1\linewidth]{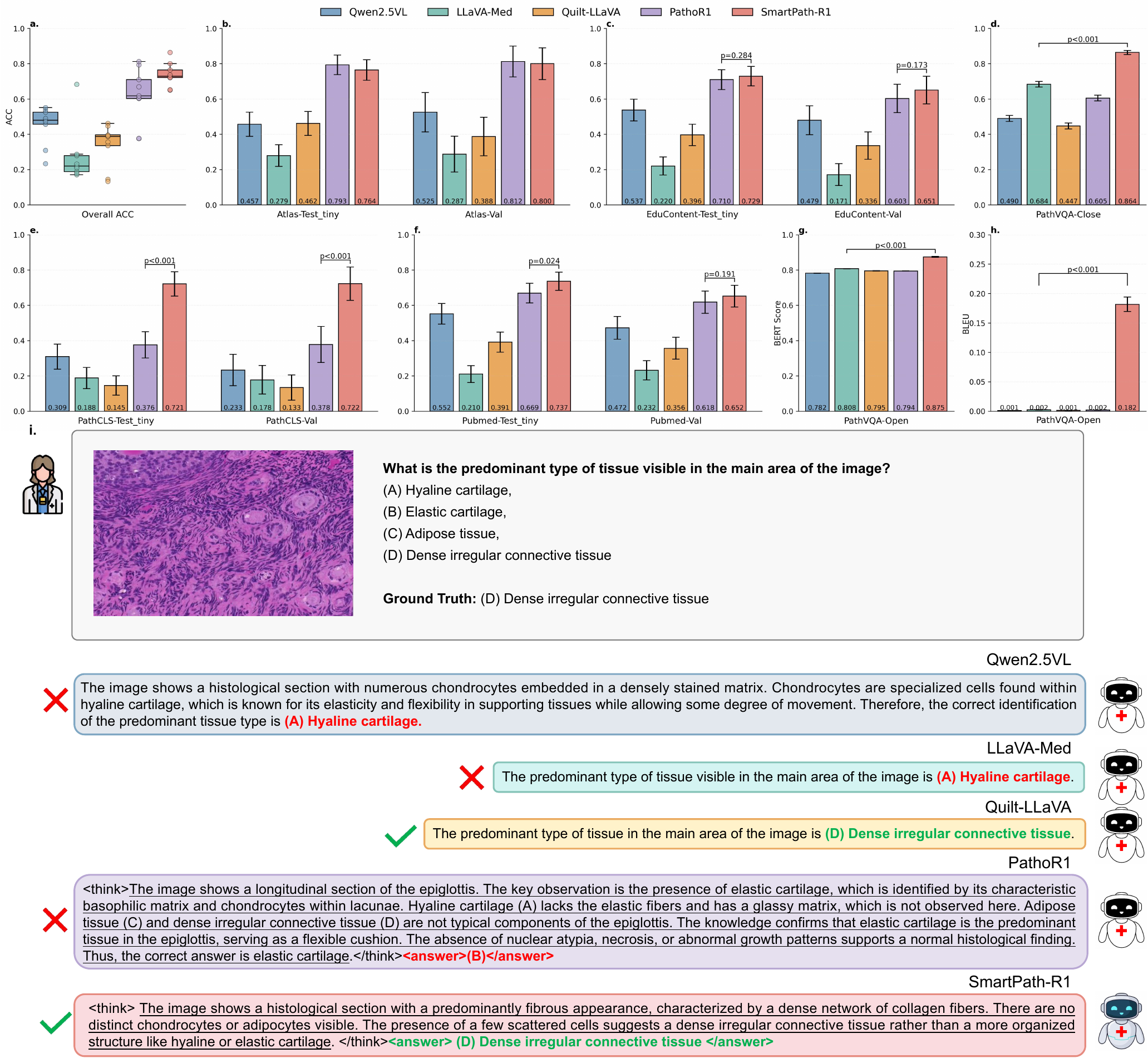}
\caption{\textbf{Performance of MLLMs on ROI-Level VQA Tasks.}
\textbf{a.} Average performance of MLLMs based on accuracy (ACC) across 10 ROI VQA tasks.
\textbf{b-h.} Model performance on specific tasks. Error bars represent 95\% CI. The box limits represent the standard error. P-values are computed using a Wilcoxon signed-rank two-sided test. Additional results are shown in Extended Data Table  \ref{tab:ROI:avg_vqa} and Table  \ref{tab:ROI:vqa}. \textbf{i.} An example of ROI-level VQA task along with the results generated by various MLLMs.}
\label{fig:ROI_vqa}
\end{figure*}

Pathological VQA is an advanced multimodal reasoning task that requires models to interpret medical images and generate accurate answers to free-form clinical questions. Unlike classification operates within a static label space (e.g., binary tumor categorization) per dataset, VQA dynamically adapts to diverse diagnostic inquiries, thereby emulating the flexible and context-driven nature of real-world clinical decision-making. This task not only demands fine-grained visual-textual alignment but also produces more interpretable and clinically actionable outputs, making it suited for complex diagnostic scenarios where reasoning and justification are critical.

To benchmark the performance of SmartPath-R1 against existing MLLMs on the fundamental ROI-level VQA task, we conduct a comparative evaluation on 10 datasets. Results demonstrate that SmartPath-R1 achieves a higher average accuracy (ACC = 0.738) compared to other models, including Qwen2.5VL (ACC = 0.450), LLaVA-Med (ACC = 0.272), Quilt-LLaVA (ACC = 0.339) and PathoR1 (ACC = 0.618) (Figures \ref{fig:ROI_vqa}b). This suggests an advantage in general ROI-level understanding. Detailed analysis of task-specific performance (Figures \ref{fig:ROI_vqa}b-h) reveals that SmartPath-R1 exhibits statistically significant (p < 0.05) improvements on tasks requiring more complex reasoning about histological structures. 

In the illustrative ROI-level VQA example (Figure \ref{fig:ROI_vqa}i), SmartPath-R1 correctly identifies the tissue type while other MLLMs fail, qualitatively demonstrating its enhanced reasoning capabilities. We attribute these gains to SmartPath-R1's training on a large-scale dataset without procedural supervision, enabling the model to learn more robust features and develop more sophisticated reasoning skills.

\subsection*{WSI-Level Classification}

WSI classification represents a transformative advancement in computational pathology, transcending the constraints of ROI paradigms to embrace the entire histological continuum. Unlike ROI-level approaches that focus on localized regions, WSI-level classification operates on the entire histological landscape, requiring the integration of multi-scale visual features with clinical knowledge to classify images. This paradigm shift from isolated ROI-level classification to holistic slide interpretation mirrors the cognitive workflow of pathologists, who routinely correlate cellular details with tissue architecture and spatial patterns to reach diagnostic conclusions. 

\begin{figure*}
    \centering
    \includegraphics[width=1\linewidth]{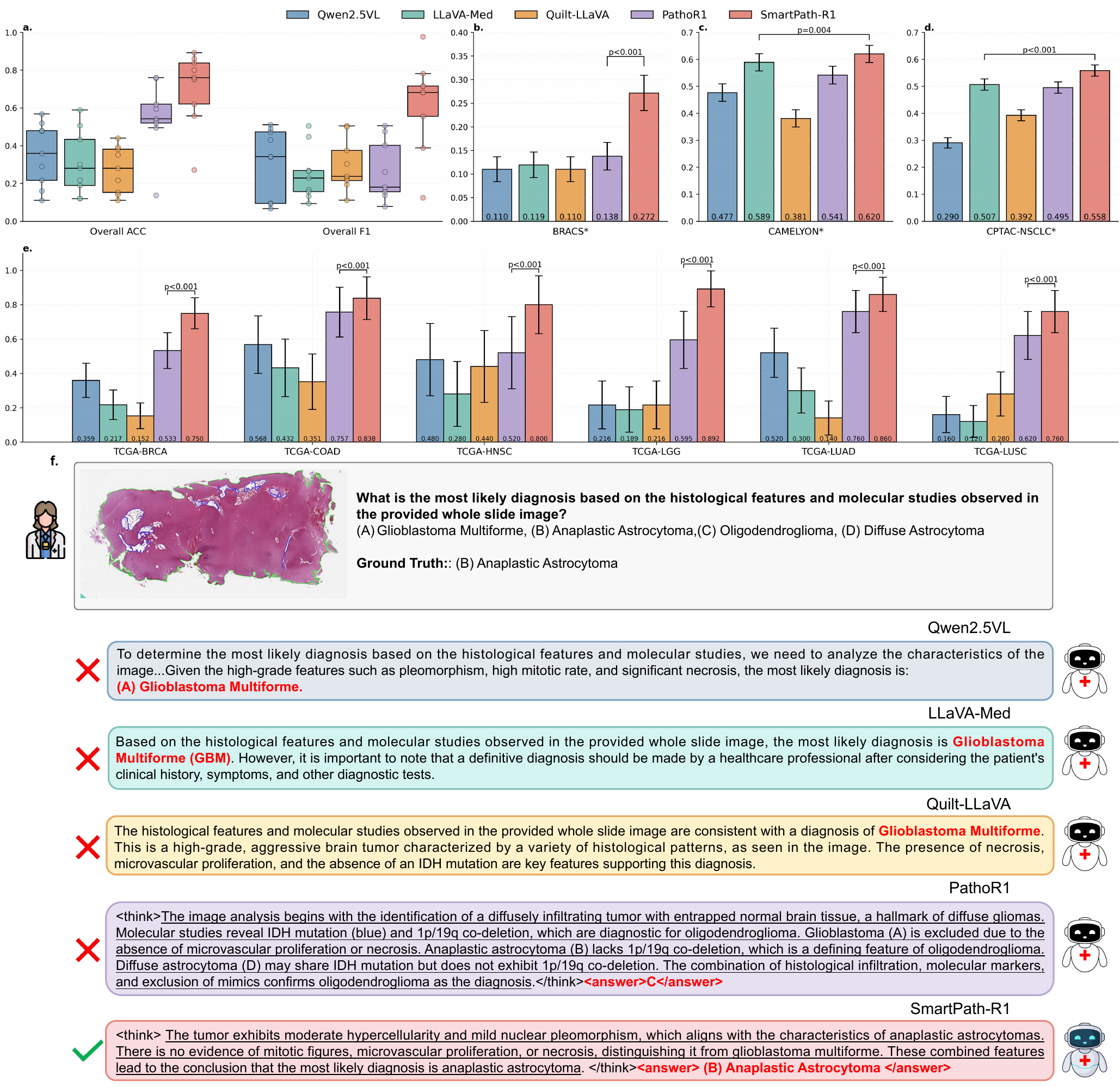}
\caption{\textbf{Performance of MLLMs on WSI-Level Classification Tasks.}
\textbf{a.} Average performance of MLLMs based on accuracy (ACC) and F1 across 9 WSI-level classification tasks.
\textbf{b-e.} Model performance on specific tasks. * represents external validation datasets. Error bars represent 95\% CI. The box limits represent the standard error. P-values are computed using a Wilcoxon signed-rank two-sided test. Additional results are shown in Extended Data Table \ref{tab:slide_cls:avg} and Table \ref{tab:WSI:tcga}. \textbf{f.} An example of WSI-level classification task along with the results generated by various MLLMs.
}
\label{fig:slide_cls}
\end{figure*}

Figure \ref{fig:slide_cls} summarizes the accuracy of five MLLMs across 9 WSI classification tasks, including 3 external ones. SmartPath-R1 consistently achieves the highest average accuracy (15.5 \% absolute margin over the second-best method; p < 0.001, Wilcoxon signed-rank test), and demonstrates superior robustness on both internal (TCGA-BRCA, TCGA-COAD, TCGA-HNSC, TCGA-LGG, TCGA-LUAD, TCGA-LUSC) and external validation sets (BRACS, CAMELYON, CPTAC-NSCLC). 

In the glioma subtyping example (Figure \ref{fig:slide_cls}f
), competing models falter due to either insufficient pathological knowledge or perceptual errors in feature recognition, leading to misclassifications. SmartPath-R1 instead performs a deliberate, step-wise exclusion: absence of mitoses, microvascular proliferation, and necrosis rules out Glioblastoma Multiforme; moderate hypercellularity and mild nuclear pleomorphism align with anaplastic astrocytoma. The model explicitly states “no evidence of… distinguishing it from glioblastoma multiforme”, yielding a coherent rationale that leads to the correct answer (B) Anaplastic Astrocytoma.

\subsection*{WSI-Level VQA}

\begin{figure*}
    \centering
    \includegraphics[width=1\linewidth]{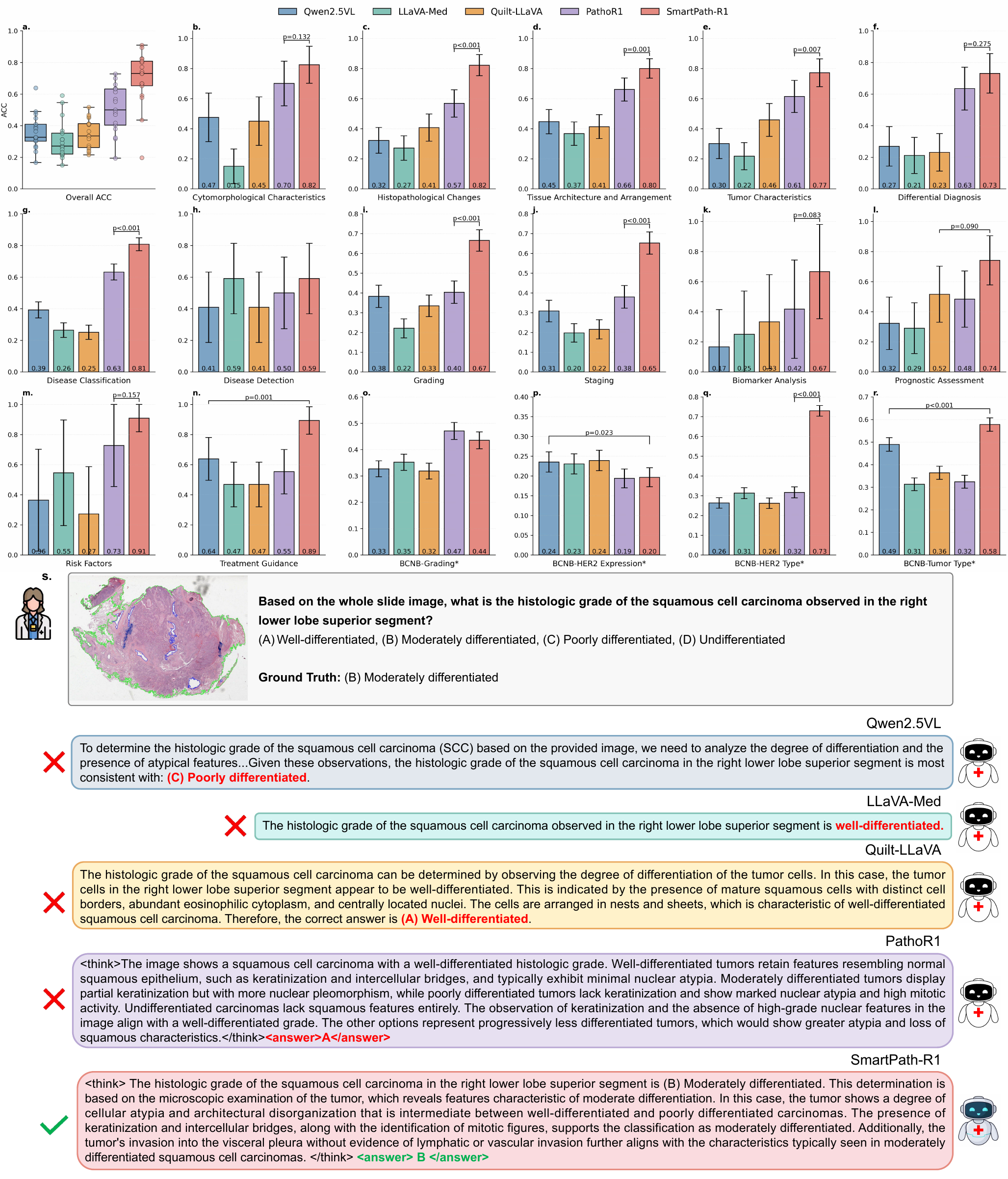}
\caption{\textbf{Performance of MLLMs on WSI-Level VQA Tasks.}
\textbf{a.} Average performance of MLLMs based on accuracy (ACC) across 13 WSI-level VQA tasks.
\textbf{b-r.} Model performance on specific tasks. * represents external validation datasets. Error bars represent 95\% CI. The box limits represent the standard error. P-values are computed using a Wilcoxon signed-rank two-sided test. Additional results are shown in Extended Data Table \ref{tab:slide_vqa:avg} and Tables \ref{tab:slide_vqa:microscopy}-\ref{tab:slide_vqa:bcnb}. \textbf{s.} An example of WSI-level VQA task along with the results generated by various MLLMs.
}
\label{fig:slide_vqa}
\end{figure*}

Unlike classification tasks, where models select from a predefined set of labels, VQA tasks introduce a more flexible and clinically aligned framework for analyzing WSIs. In WSI-level VQA, models process pathology relevant questions about gigapixel WSIs and select from dynamic, context-dependent options. Accurately answering questions about WSIs requires the model to understand the global context, integrate information from different regions, apply domain knowledge, and handle ambiguity.

To evaluate the performance of SmartPath-R1 at the WSI level, we conduct a comparative analysis against existing MLLMs on 13 WSI-level VQA tasks (Figure \ref{fig:slide_vqa}). 
The results demonstrate that SmartPath-R1 achieves a significantly higher average ACC (0.695) across all tasks (Figures \ref{fig:slide_vqa}a) compared to other models (PathoR1: 0.505, Quilt-LLaVA: 0.350, LLaVA-Med: 0.309, Qwen2.5VL: 0.360), indicating superior overall performance. This suggests that SmartPath-R1 possesses a greater capacity for reasoning about complex histological patterns and integrating information across the entire WSI.
Analysis of task-specific performance (Figures \ref{fig:slide_vqa}b-r) reveals that SmartPath-R1 exhibits particularly strong performance on questions related to histopathological changes, disease classification, grading, staging, and treatment guidance (p < 0.001), suggesting its ability to reason about tumor-specific features in WSIs. 

The comparative evaluation further presents SmartPath-R1's critical advantage in reconciling diagnostic ambiguity, as exemplified in the squamous cell carcinoma (SCC) grading task (Figure \ref{fig:slide_vqa}s), where it correctly identifies the moderately differentiated (B) ground truth diagnosis. While other models produce binary interpretations (either well- or poorly-differentiated), SmartPath-R1 recognizes the nuanced histologic spectrum for moderate differentiation.

\section{Discussion}

The development of SmartPath-R1 represents a paradigm shift in computational pathology by addressing two critical limitations of existing MLLMs: (1) the lack of clinically meaningful reasoning capacity and (2) the artificial fragmentation of tasks across biological scales. Our results demonstrate that embedding sequential reasoning and multiscale integration into MLLMs significantly enhances diagnostic accuracy, interpretability, and clinical relevance.

\noindent \textbf{Reasoning as a Cornerstone of Diagnostic AI.}
Traditional supervised fine-tuning approaches minimize prediction errors but fail to capture the sequential decision-making process that pathologists employ. SmartPath-R1 overcomes this limitation by framing pathology interpretation as a reinforcement learning problem, where the model learns optimal evidence-gathering policies through reward signals. This approach allows the model to progressively integrate visual cues with language context, explicitly rule out competing diagnoses, and provide transparent justifications via stepwise \emph{<think></think><answer></answer>} reasoning chains, mirroring clinical decision-making.
Our benchmarking against state-of-the-art MLLMs (Qwen2.5VL, LLaVA-Med, Quilt-LLaVA, and PathoR1) reveal that models lacking structured pathological reasoning frequently misclassify diagnostically challenging cases. The superior performance of SmartPath-R1 underscores that diagnostic accuracy in pathology depends not just on visual recognition but on the ability to synthesize evidence hierarchically.

\noindent \textbf{Multiscale Integration for Holistic Pathology Analysis.}
Clinical pathology requires simultaneous evaluation of subcellular details (ROI-level) and architectural patterns (WSI-level), which is absent in current MLLMs. SmartPath-R1 bridges this gap via a dynamic mixture-of-experts mechanism, enabling adaptive switching of ROI-level morphometrics and WSI-level topological reasoning.
This architecture eliminates the need for separate ROI-based and WSI models, streamlining workflows while improving diagnostic consistency. 

\noindent \textbf{Clinical Implications.}
SmartPath-R1's reinforcement learning based training paradigm offers transformative advantages over conventional supervised learning approaches, significantly advancing clinical applicability. By learning reasoning policies directly from endpoint labels, the model substantially reduces dependency on labor-intensive procedural annotations while more accurately capturing the dynamic decision-making process of pathologists. 
This approach enhances scalability and adaptability to various cancer subtypes, where limited training data traditionally constrain AI performance.
Furthermore, the built-in explainability of SmartPath-R1's stepwise reasoning process fosters clinician trust, a pivotal factor for real-world adoption in diagnostic workflows. Together, these capabilities position SmartPath-R1 as a robust and scalable solution for next-generation computational pathology.

\noindent \textbf{Future Work.}
Future research should focus on advancing AI-powered pathology diagnostics by addressing current limitations while expanding the system's capabilities. A critical direction involves integrating multimodal data beyond image and language, such as molecular profiles and clinical records, to align with modern precision pathology workflows. This expansion requires robust fusion architectures to harmonize diverse data types while mitigating potential biases introduced by heterogeneous sources.
To enhance the model's clinical utility, efforts should prioritize improving its reasoning transparency and adaptability. For example, incorporating retrieval-augmented generation (RAG) would allow the system to dynamically reference medical literature and case databases, grounding its predictions in evidence. 

In conclusion, SmartPath-R1 establishes a new standard for AI in pathology by demonstrating that human-like diagnostic reasoning can be computationally modeled and optimized. By unifying multiscale analysis with reinforcement learning-driven decision-making, our framework transcends the limitations of current MLLMs, offering a scalable, interpretable, and clinically actionable solution. This work not only advances computational pathology but also provides a blueprint for developing next-generation diagnostic AI across medicine, where reasoning, not just recognition, determines success.

\section{Methods}

\subsection{Model Architecture}

Our model is built upon the Qwen2.5-VL \cite{bai2025qwen2} architecture, which comprises three primary components: a Large Language Model (LLM), a Vision Encoder, and an MLP-based Vision-Language Merger. The LLM is initialized with pre-trained weights from the Qwen2.5 LLM and incorporates a modified 1D Rotary Position Embedding (RoPE) to better handle multimodal inputs. The Vision Encoder utilizes a redesigned Vision Transformer (ViT) architecture with 2D-RoPE and window attention to support native input resolutions and accelerate computation. Input images are resized to multiples of 28 and processed by splitting them into patches. Finally, an MLP-based Vision-Language Merger addresses the efficiency challenges of long image feature sequences by spatially grouping adjacent sets of four patch features, concatenating them, and projecting them into a lower-dimensional space that aligns with the text embeddings. 

To enhance the model's adaptability and efficiency, we employ Low-Rank Adaptation (LoRA) \cite{hu2022lora} for different tasks. LoRA allows for efficient adaptation of the pre-trained MLLM to specific tasks by training low-rank matrices that represent parameter updates, while keeping the original MLLM weights frozen. This approach significantly reduces the number of trainable parameters and computational cost compared to full fine-tuning. To leverage the strengths of different LoRA configurations, we adopt a Mixture of Experts (MoE) \cite{jordan1994hierarchical,jacobs1991adaptive} approach. Specifically, we train multiple LoRA modules, each tailored to a specific group of tasks. During inference, a routing mechanism selects the most appropriate LoRA module based on the input language prompt. This allows the model to specialize in different aspects of the task while maintaining a shared foundation in the pre-trained MLLM.

\subsection{Details of Model Training}

\subsubsection*{Scale-dependent Supervised Fine-Tuning}

To address the computational challenges associated with high-resolution histopathology images, we introduce a scale-dependent supervised fine-tuning strategy. Recognizing that different tasks require varying levels of image detail, we dynamically adjust the image resolution and resulting token count based on the task's inherent scale. Specifically, for ROI-level tasks, images are transformed to a lower resolution to achieve a smaller token count, thereby reducing computational burden.
For WSI-level tasks, which require a broader contextual understanding of the tissue architecture, images are transformed to a higher resolution, resulting in a larger token count to preserve critical contextual information.

This scale-dependent resizing is achieved using the transformation function ${T}^{M,P}: \mathbb{R}^{H \times W \times 3} \rightarrow \mathbb{R}^{H' \times W' \times 3}$. Given an input image of height $H$ and width $W$, the transformation function outputs a resized image with height $H'$ and width $W'$. The parameters $M$ and $P$ control the maximum token count and patch size, respectively. Specifically, $M$ is set to 256 for ROI-level tasks and 1024 for WSI-level tasks. The transformation ensures that the output dimensions $H'$ and $W'$ are multiples of the patch size $P$ and that the total number of tokens $H'W'/ P^2$ does not exceed $M$. This is achieved by computing a scaling factor and applying $P$-aligned rounding to the original dimensions.

The supervised fine-tuning process is then formalized as follows:
Given a dataset $\mathcal{D} = {(\mathbf{I}_i, \mathbf{y}_i)}_{i=1}^{|\mathcal{D}|}$, where $\mathbf{I}_i$ represents the transformed image (either ROI-level or WSI-level, depending on the task) and $\mathbf{y}_i$ represents the corresponding ground truth answer, the objective is to minimize the cross-entropy loss:
\begin{equation} 
\mathcal{L}_{SFT} = - \frac{1}{|\mathcal{D}|} \sum_{i=1}^{|\mathcal{D}|} \log p(\mathbf{y}_i | \mathbf{I}_i; \theta), 
\end{equation}
where $p(\mathbf{y}_i | \mathbf{I}_i; \theta)$is the model's predicted probability of the correct label $\mathbf{y}_i$ given the input image $\mathbf{I}_i$ and model parameters $\theta$. The model parameters $\theta$ are then updated using gradient descent to minimize $\mathcal{L}_{SFT}$.

\subsubsection*{Group Relative Policy Optimization}
Group relative policy optimization (GRPO) \cite{deepseekr1} is an efficient reinforcement learning algorithm derived from Proximal Policy Optimization (PPO) \cite{schulman2017proximal}. It eliminates the need for a separate critic model by estimating the baseline value directly from group statistics, thereby substantially reducing computational requirements while maintaining stable policy updates. For each question $q$ and pathology image $\mathbf{I}$, GRPO samples a group of outputs $\{o_1, o_2, \cdots, o_G\}$ from the old policy $\pi_{\theta_{old}}$ and then optimizes the policy model $\pi_{\theta}$ by maximizing the following objective:
\begin{equation}
\begin{split}
    \mathcal{J}&_{GRPO}(\theta) = \mathbb{E}{[q,\mathbf{I} \sim P, \{o_g\}_{g=1}^G \sim \pi_{\theta_{old}}(O|q,\mathbf{I})]}  \\
    & \frac{1}{G}\sum_{g=1}^G \left( \min \left( \frac{\pi_\theta(o_g |q,\mathbf{I})}{\pi_{\theta_{old}}(o_g |q,\mathbf{I})} A_g, \text{clip} \left( \frac{\pi_\theta(o_g |q,\mathbf{I})}{\pi_{\theta_{old}}(o_g |q,\mathbf{I})}, 1 - \epsilon, 1 + \epsilon \right)  A_g \right) - \beta \mathbb{D}_{KL}\left(\pi_{\theta} || \pi_{ref}\right)\right) ,
\end{split}
\label{eq:GRPO-obj}
\end{equation}
\begin{equation}
    \mathbb{D}_{KL}\left(\pi_{\theta} || \pi_{ref}\right) = \frac{\pi_{ref}(o_g|q,\mathbf{I})}{\pi_{\theta}(o_g|q,\mathbf{I})}- \log\frac{\pi_{ref}(o_g|q,\mathbf{I})}{\pi_{\theta}(o_g|q,\mathbf{I})} - 1,
\end{equation}
where $\epsilon$ and $\beta$ are hyper-parameters, and $A_i$ is the advantage, computed using a group of rewards $\{r_1, r_2, \ldots, r_G\}$ corresponding to the outputs within each group:
\begin{equation}
    A_g = \frac{r_g - {\operatorname{mean}(\{r_1, r_2, \cdots, r_G\})}}{{\operatorname{std}(\{r_1, r_2, \cdots, r_G\})}}.
\end{equation}

This formulation provides several key advantages: (1) The group-based advantage estimation eliminates the need for a separate value network, reducing model complexity; (2) The relative reward normalization within each group ensures stable policy updates across varying reward scales; (3) The combined clipping and KL penalty mechanisms prevent excessively large policy updates while maintaining sufficient exploration capacity.

\begin{table}[t]
\caption{\textbf{Pathological Reasoning Templates for Different Tasks.}}
\label{tab:prompts}
\begin{center}
{
\begin{tabular}{p{13.5cm}}  
\toprule
\textbf{System Prompt $\mathsf{p^{sys}}$:} A conversation between User and Assistant. The user asks a question, and the assistant solves it. The assistant first thinks about the reasoning process in the mind and then provides the user with the answer. The reasoning process and answer are enclosed within <think> </think> and <answer> </answer> tags, respectively, i.e., <think> reasoning process here </think>  <answer> answer here </answer>.\\
\midrule
\textbf{Classification Prompt $\mathsf{p^{cls}}$:} Classify this pathological image into one of these classes: (A) \emph{\{Class\_A\}}, (B) \emph{\{Class\_B\}}, (C) \emph{\{Class\_C\}}...\\
\midrule
\textbf{VQA Prompt $\mathsf{p^{vqa}}$:} \emph{\{user\_question\}}?\\
\midrule
\textbf{Detection Prompt $\mathsf{p^{det}}$:} Detect \emph{\{pathological\_category\}} in pathology \emph{\{anatomical region\}}. Output bounding boxes in [[x\_min, y\_min, x\_max, y\_max],...] format.\\
\midrule
\textbf{Segmentation Prompt $\mathsf{p^{seg}}$:} Segment \emph{\{pathological\_category\}} in pathology \emph{\{anatomical region\}}. Output bounding boxes in [[x\_min, y\_min, x\_max, y\_max],...] format.\\
\bottomrule
\end{tabular}
}
\end{center}
\end{table}

\subsubsection*{Task-Aware Reinforcement Fine-Tuning}
We design specialized reward functions for different critical pathological tasks: classification, VQA, detection, and segmentation. As shown in Table  \ref{tab:prompts}, each task utilizes distinct pathological reasoning prompts: a system prompt $\mathsf{p^{sys}}$, and task-specific prompts ($\mathsf{p^{cls}}$,$\mathsf{p^{vqa}}$, $\mathsf{p^{det}}$, $\mathsf{p^{seg}}$). The model generates structured responses containing both the reasoning process and final answer, which are parsed to extract task-specific outputs.
All reward functions combine two key components:
\begin{equation}
R = R_{\text{task}} + \lambda R_{\text{format}},
\end{equation}
where $R_{\text{task}}$ evaluates task performance and $R_{\text{format}}$ enforces response standardization (<think></think> for reasoning and <answer></answer> for final outputs). The weighting parameter $\lambda$ is set to 1 for balanced optimization.

\noindent \textbf{Classification Reward.}
For a given histopathology image $\mathbf{I}$ and question, the model produces $\mathsf{r} = \operatorname{TP}(\mathbf{I}, \mathsf{p^{sys}}, \mathsf{p^{cls}})$ with predicted subtype $\hat{\mathsf{y}}$. The reward function:
\begin{equation}
R_{\text{cls}} = \mathbb{I}(\hat{\mathsf{y}} = \mathsf{y}) + \lambda R_{\text{format}}(\mathsf{r}),
\end{equation}
where $\mathsf{y} \in \mathcal{Y}$ is the ground truth label from the possible subtypes $\mathcal{Y}$. This binary reward structure emphasizes precise diagnostic classification.

\noindent \textbf{Detection Reward.}
For detection tasks, the model outputs $\mathsf{r} = \operatorname{TP}(\mathbf{I}, \mathsf{p^{sys}}, \mathsf{p^{det}})$ with parsed bounding boxes $\mathcal{P} = \{\mathbf{p_i}\}_{i=1}^I$. The reward combines:
\begin{equation}
R_{\text{det}} = \text{AP50}(\mathcal{P}, \mathcal{G}) + \lambda R_{\text{format}}(\mathsf{r}),
\end{equation}
where $\mathcal{G} = \{\mathbf{g_j}\}_{j=1}^J$ are ground truth annotations and $\text{AP50}$ computes average precision at 0.5 intersection-over-union threshold. This formulation jointly optimizes for both localization accuracy and standardized output format.

\noindent \textbf{Segmentation Reward.}
Since MLLMs struggle to directly output pixel-level segmentation masks, we first provide spatial prompts in the form of bounding box coordinates, then leverage an off-the-shelf segmentation model, MedSAM \cite{ma2024segment}, to obtain the segmentation results.  Specifically, the model outputs $\mathsf{r} = \operatorname{TP}(\mathbf{I}, \mathsf{p^{sys}}, \mathsf{p^{seg}})$ with parsed bounding boxes $\mathcal{P} = \{\mathbf{p_i}\}_{i=1}^I$. The reward combines:
\begin{equation}
R_{\text{seg}} = \text{Dice}(\text{MedSAM}(\mathcal{P}), \mathbf{M}) + \lambda R_{\text{format}}(\mathsf{r}),
\end{equation}
where $\mathbf{M}$ are ground truth annotations. This formulation jointly optimizes for both segmentation accuracy and standardized output format.

\noindent \textbf{VQA Reward.}
For VQA task, the model generates response $\mathsf{r} = \operatorname{TP}(\mathbf{I}, \mathsf{p^{sys}}, \mathsf{p^{vqa}})$ with parsed answer $\hat{\mathsf{a}}$. The composite reward incorporates answer accuracy and format compliance:
\begin{equation}
R_{\text{vqa}} = R_{\text{ans}}(\hat{\mathsf{a}}, \mathsf{a}) + \lambda R_{\text{format}}(\mathsf{r}).
\end{equation}
The answer reward differs by question type:
\begin{equation}
R_{\text{ans}}(\hat{\mathsf{a}}, \mathsf{a}) =
\begin{cases}
\mathbb{I}(\hat{\mathsf{a}} = \mathsf{a}), & \text{(closed-ended)} \\
\text{BLEU-4}(\hat{\mathsf{a}}, \mathsf{a}), & \text{(open-ended)}
\end{cases}
\end{equation}
where $\mathsf{a}$ denotes ground truth, $\mathbb{I}(\cdot)$ is the indicator function, and BLEU-4 evaluates textual similarity for open-ended responses.

The unified reward architecture enables simultaneous optimization of clinical accuracy and interpretability, while task-specific components ensure appropriate evaluation metrics for each diagnostic modality. The format compliance term ($R_{\text{format}}$) consistently enforces structured reasoning outputs across all tasks, facilitating clinical validation of model decisions.

\subsection{Curated Datasets}

This study leverages a diverse collection of histopathology datasets, categorized into ROI-level image-text pairs, ROI-level classification, ROI-level detection and segmentation, ROI-level VQA, WSI-level image-text pairs, WSI-level classification, and WSI-level VQA. Specifically, PathCap, PathInstruct, Quilt-1M, and SlideInstruction-Caption are used for supervised fine-tuning. The training sets from CCRCC, Chaoyang, CRC-MSI, ESCA, Pancancer-TIL, UniToPatho (ROI-level classification, see Extended Data Table \ref{tab:ROI:classes_cls} for classes), BCSS, CoCaHis, CoNIC2022, CRAG, DigestPath2019 Cell, DigestPath2019 Tissue, MIDOG Breast, MIDOG Neuroendocrine, Rings, SICAPv2, TNBC, WSS4LUAD (ROI-level detection and segmentation, see Extended Data Table \ref{tab:ROI:classes_det} for classes), PathMMU, PathVQA (ROI-level VQA), TCGA (WSI-Level classification, see Extended Data Table \ref{tab:WSI:classes_cls} for classes), and SlideInstruction-VQA (WSI-Level VQA) are used for supervised fine-tuning and reinforcement fine-tuning. BreakHis, PanCancer-TCGA, GlaS, NuCLS, BRACS, CAMELYON, CPTAC-NSCLC, and BCNB are exclusively used for external validation to assess model generalizability. Importantly, to prevent potential data leakage, we meticulously screened and removed any samples from the original test sets that might have been inadvertently used during training (including potential overlaps across different groups of tasks). This precaution maintains the integrity of our validation process.

\subsubsection*{ROI-Level Image-Text Pairs}

\textbf{PathCap.}
The PathCap dataset \cite{pathasst} is a curated collection of 207,000 high-quality pathology image-caption pairs. The majority of the data (197,000 pairs) are sourced from PubMed publications and internal pathology guideline books, followed by rigorous filtering to ensure pathological relevance. An additional 10,000 expert-annotated pairs are contributed by cytologists specializing in liquid-based cytology (LBC), enhancing the dataset's clinical precision. To address challenges such as non-pathological noise and suboptimal image clarity in raw PubMed data, the authors employ a multi-stage refinement pipeline: (1) A ConvNeXt-based classifier trained on 20,000 manually annotated samples filters out non-pathological images, yielding 135,000 pathology-specific candidates; (2) A YOLOv7 model segments composite figures into sub-images, while ChatGPT separates and aligns intricate captions, supported by PLIP for visual-textual similarity assessment; (3) Captions are refined via ChatGPT to remove extraneous clinical metadata (e.g., patient age) and standardize descriptive styles. The resulting dataset offers a diverse, high-resolution resource for training vision-language models in pathology, with explicit emphasis on diagnostic relevance.

\noindent \textbf{PathInstruct.}
The PathInstruct dataset \cite{pathasst} comprises 180,000 multimodal instruction-following samples tailored to advance interactive AI applications in pathology. It is structured into two components: (1) ChatGPT-generated instructions derived from curated pathology image-text pairs (those with captions exceeding 12 words), which include both detailed descriptions (e.g., inquiries about histological features) and conversational Q\&A formats mimicking clinician-AI dialogues; and (2) Specialized model-invoking instructions that guide AI systems to dynamically leverage pathology-specific sub-models based on user intent and image features. The dataset construction leverages the refined outputs of PathCap, with additional prompts engineered to elicit task-oriented responses.

\noindent \textbf{Quilt-1M.}
The Quilt-1M dataset \cite{ikezogwo2023quilt} represents a significant advancement in large-scale vision-language data for histopathology, offering one million meticulously aligned image-text pairs to support multimodal AI research. Derived from an initial collection of 437,878 images and 802,144 text descriptions extracted from 1,087 hours of expert-curated educational YouTube videos, the dataset spans multiple microscopic magnification levels to ensure comprehensive pathological coverage. The construction process leverages a combination of LLM (GPT-3.5), automatic speech recognition, and human-validated algorithms to accurately pair histopathology images with relevant textual descriptions. To further enhance diversity and scale, the dataset incorporates supplementary data from sources such as Twitter, research publications, and web-crawled resources, while deliberately avoiding overlap with existing open-access repositories.

\subsubsection*{ROI-Level Classification}
\textbf{CCRCC (4 classes).}
The CCRCC dataset \cite{brummer2023computational} comprises 52,723 annotated histopathology ROIs (300$\times$300 pixels) derived from WSIs of clear cell renal cell carcinoma (CCRCC) specimens. These ROIs are randomly sampled from two independent sources: the TCGA-KIRC repository and the Helsinki cohort.
The dataset encompasses six distinct histological classes: malignant tumor regions (13,057 ROIs), normal renal parenchyma (8,652 ROIs), stromal tissue (5,460 ROIs), red blood cell accumulations (996 ROIs), non-informative background areas (16,026 ROIs), and heterogeneous tissue types including necrosis and artifacts (8,522 ROIs). For robust classification modeling, we focused exclusively on four biologically meaningful classes - cancer, normal tissue, stroma, and blood - excluding ambiguous and non-informative ROIs.
The dataset is randomly partitioned into training (22,530 ROIs) and test (5,635 ROIs) sets while preserving class distributions. This curated subset facilitates precise evaluation of computational pathology algorithms while minimizing confounding factors from ambiguous labels. Comprehensive performance metrics are provided in Extended Data Table \ref{tab:ROI:CCRCC}.

\noindent \textbf{Chaoyang (4 classes).} 
The Chaoyang dataset \cite{zhu2021hard} provides a comprehensive collection of histopathology ROIs for colorectal tissue analysis, comprising four clinically relevant classes: normal mucosa (1,816 ROIs), serrated lesions (1,163 ROIs), adenocarcinoma (2,244 ROIs), and adenoma (937 ROIs). All ROIs are standardized to 224$\times$224 pixels and divided into training (4,021 ROIs) and test (2,139 ROIs) sets using the official split to ensure reproducibility.
Performance metrics are detailed in Extended Data Table \ref{tab:ROI:Chaoyang}.

\noindent \textbf{CRC-MSI (2 classes).} 
The CRC-MSI dataset comprises 51,918 high-resolution histopathology ROIs (512$\times$512 pixels) derived from colorectal cancer specimens in the TCGA database \cite{kather2019deep}. Each ROI is annotated with patient-level Microsatellite Instability (MSI) status, categorized as either MSI-H (Microsatellite Instability-High)
or NonMSIH (combining Microsatellite Instability-Low and Microsatellite Stable cases). Using the official dataset partition, we maintained rigorous separation between training (19,557 ROIs) and test (32,361 ROIs) sets to ensure unbiased evaluation. This binary classification framework enables the development of AI models for MSI status prediction directly from histopathology images, a clinically important molecular characteristic in colorectal cancer.
Performance metrics are reported in Extended Data Table \ref{tab:ROI:CRCMSI}.

\noindent \textbf{ESCA (11 classes).} 
The ESCA dataset \cite{tolkach2023artificial} comprises 367,229 histopathology ROIs (256$\times$256 pixels) extracted from 320 WSIs of esophageal adenocarcinoma and esophagogastric junction tumors, collected from four institutions: University Hospital Cologne (UKK, 22 slides), Landesklinikum Wiener Neustadt (WNS, 62 slides), TCGA (22 slides), and Charité - Universitätsmedizin Berlin (CHA, 214 slides). Each ROI is annotated with one of eleven histological classes: adventitia (71,131 ROIs), lamina propria mucosae (2,173 ROIs), muscularis mucosae (2,951 ROIs), muscularis propria (83,358 ROIs), regression tissue (56,490 ROIs), gastric mucosa (44,416 ROIs), esophageal mucosa (18,561 ROIs), submucosa (22,117 ROIs), submucosal glands (1,516 ROIs), tumor (63,863 ROIs), and ulceration (753 ROIs).
For model development, we used the CHA dataset (178,187 ROIs) for training and combined UKK, WNS, and TCGA datasets (189,142 ROIs) for testing. All ROIs are resized to 224$\times$224 pixels for consistency. This multi-center dataset enables robust evaluation of AI models for esophageal carcinoma subtyping, with results reported in Extended Data Table \ref{tab:ROI:ESCA}. 

\noindent \textbf{Pancancer-TIL (2 classes).}
We utilize the PanCancer-TIL dataset \cite{abousamra2022deep, saltz2018spatial} containing 304,097 histopathology ROIs (100$\times$100 pixels at 0.5 micrometers per pixel) for tumor-infiltrating lymphocyte classification. The dataset includes 54,910 TIL-positive ROIs (at least two TILs present in the image) and 249,190 TIL-negative ROIs, following the official train-val-test split (209,221:38,601:56,275 ROIs). All ROIs are resized to 256$\times$256 pixels for model input standardization. Performance metrics are reported in Extended Data Table \ref{tab:ROI:PanCancerTIL}.

\noindent \textbf{UniToPatho (6 classes).} 
The UniToPatho dataset \cite{barbano2021unitopatho} provides 9,536 histopathology ROIs extracted from 292 WSIs of colorectal specimens, specifically annotated to support deep learning-based classification of colorectal polyps and adenoma grading. The dataset includes six clinically relevant classes: Normal tissue (950 ROIs), Hyperplastic Polyp (545 ROIs), Tubular Adenoma with High-Grade dysplasia (454 ROIs), Tubular Adenoma with Low-Grade dysplasia (3,618 ROIs), Tubulo-Villous Adenoma with High-Grade dysplasia (916 ROIs), and Tubulo-Villous Adenoma with Low-Grade dysplasia (2,186 ROIs).
Following the official dataset partition, we used 6,270 ROIs for training and 2,399 ROIs for testing. This carefully curated collection enables the development of AI models for precise polyp characterization, with experimental results detailed in Extended Data Table \ref{tab:ROI:UniToPatho}. 

\noindent \textbf{BreakHis* (2 classes).}
The BreakHis dataset \cite{spanhol2015dataset} for breast cancer histopathological image classification is utilized in this study for external validation. The dataset comprises two primary classes: benign tumors and malignant tumors. All ROIs are captured at four distinct magnification levels (40$\times$, 100$\times$, 200$\times$, and 400$\times$). For consistency, images are resized to 224$\times$224 pixels. For external validation, only the test set is employed, consisting of 1,582 ROIs (20\% of the total dataset), with stratification to preserve the original label distribution. Detailed experimental results are provided in Extended Data Table \ref{tab:ROI:BreakHis}.

\noindent \textbf{PanCancer-TCGA* (32 classes).}
The PanCancer-TCGA dataset \cite{komura2022universal} is also employed for external validation in this study. We utilize 2,000 histopathology images across 32 distinct cancer types, including Head and Neck Squamous Cell Carcinoma, Bladder Urothelial Carcinoma, Uterine Carcinosarcoma, Colon Adenocarcinoma, Lymphoid Neoplasm Diffuse Large B-cell Lymphoma, Lung Squamous Cell Carcinoma, Brain Lower Grade Glioma, Esophageal Carcinoma, Pheochromocytoma and Paraganglioma, Sarcoma, Glioblastoma Multiforme, Adrenocortical Carcinoma, Uterine Corpus Endometrial Carcinoma, Prostate Adenocarcinoma, Breast Invasive Carcinoma, Stomach Adenocarcinoma, Pancreatic Adenocarcinoma, Skin Cutaneous Melanoma, Ovarian Serous Cystadenocarcinoma, Thymoma, Lung Adenocarcinoma, Kidney Renal Papillary Cell Carcinoma, Testicular Germ Cell Tumors, Kidney Renal Clear Cell Carcinoma, Rectum Adenocarcinoma, Cholangiocarcinoma, Cervical Squamous Cell Carcinoma and Endocervical Adenocarcinoma, Thyroid Carcinoma, Mesothelioma, Uveal Melanoma, Liver Hepatocellular Carcinoma, and Kidney Chromophobe. The test set is rigorously stratified to maintain class distribution, ensuring an unbiased assessment of model generalizability. Performance metrics, demonstrating superior results compared to baseline models, are detailed in Extended Data Table \ref{tab:ROI:PanCancerTCGA}.

\subsubsection*{ROI-Level Detection and Segmentation}

\textbf{BCSS.} 
The Breast Cancer Semantic Segmentation (BCSS) dataset \cite{amgad2019structured} is a comprehensive histopathology image dataset for breast cancer analysis derived from TCGA WSIs and annotated by expert pathologists through the Digital Slide Archive platform. It contains 9,192 annotated images uniformly resized to 1024$\times$1024 pixels, including 7,322 training images and 1,870 test images. The dataset provides detailed pixel-level annotations across 16 distinct tissue classes essential for pathological diagnosis: Mucoid material, Blood, Metaplasia NOS, Glandular secretions, Necrosis or debris, Plasma cells, Lymphatics, Blood vessel, Other immune, Dcis, Tumor, Stroma, Normal acinus or duct, Fat, Lymphocytic, and Angioinvasion.

\noindent \textbf{CoCaHis.} 
CoCaHis comprises 82 hematoxylin-eosin stained frozen section images acquired intraoperatively from 19 patients with liver metastases of colon cancer. The dataset includes pixel-wise annotations created through a consensus process involving four board-certified pathologists, two pathology residents, and one final-year medical student, demonstrating substantial inter-rater agreement. Following the official data partitioning, we employed 58 images for training and 24 for testing, with all images standardized to 1024$\times$1024 pixels. 

\noindent \textbf{CoNIC2022.}
The CoNIC2022 dataset \cite{graham2024conic} is utilized for nuclear segmentation in colorectal pathology collected across 16 institutions in three countries, with nuclei classified into six classes: epithelial cells, lymphocytes, plasma cells, neutrophils, eosinophils, and connective tissue cells (a composite class encompassing endothelial cells, fibroblasts, and muscle cells). 
Segmentation masks are generated using a semi-automated approach with subsequent manual refinement, yielding a combined dataset of 535,063 nuclei. While this hybrid annotation strategy enabled large-scale data curation, we acknowledge the potential for residual noise inherent in semi-automated methods, which we systematically evaluated through comparative pathologist review.
All image ROIs are standardized to 1024$\times$1024 pixels and partitioned into 15,387 training images and 3,945 test images.

\noindent \textbf{CRAG.}
The CRAG dataset \cite{graham2019mild} contains 213 H\&E colorectal adenocarcinoma WSIs for gland segmentation at 20x magnification from the University Hospitals Coventry and Warwickshire (UHCW) NHS Trust in Coventry, United Kingdom. We divide the slices into 1024$\times$1024 ROIs, resulting in 1,429 and 321 images for training and testing, respectively.

\noindent \textbf{DigestPath2019 Cell.} 
The DigestPath2019 Cell dataset \cite{da2022digestpath} is clinically validated for signet ring cell carcinoma (SRCC) detection and segmentation, representing the first public resource specifically designed for this diagnostically challenging gastrointestinal cancer variant. The dataset is collected from 155 patients across gastric and intestinal mucosa, sourced from four leading Chinese medical institutions to ensure geographic and demographic diversity. Annotations underwent rigorous multi-tiered review by gastrointestinal pathologists, with senior experts adjudicating discordant cases to ensure label accuracy. All images are standardized to 1024$\times$1024 pixels to balance computational efficiency with diagnostic fidelity. The dataset is partitioned into 352 training and 85 testing images.

\noindent \textbf{DigestPath2019 Tissue.} 
The DigestPath2019 Tissue dataset \cite{da2022digestpath} consists of total 872 tissue sub-slices from 476 patients, which are extracted from both benign and malignant areas to cover as much variety of tissue appearance as possible. Two expert pathologists review the sub-images to ensure no uncertain tissues between benign and malignant. Then, the malignant images are manually annotated at pixel-level by four experienced pathologists, followed by expert pathologist examination. We divide the slices into 1024$\times$1024 images, resulting 10,725 and 2,666 images for training and testing, respectively.

\noindent \textbf{MIDOG Breast.}
The MIDOG Breast dataset \cite{aubreville2024domain} comprises 150 WSI cases of human breast carcinoma evenly distributed across three digital scanning systems at 40$\times$ magnification obtained from the UMC Utrecht pathology archive. For computational analysis, these cases are processed into 3,794 annotated image ROIs, divided into a training set (3,032 ROIs) and an independent test set (762 ROIs). Each instance can be classified into one of two classes: breast cancer mitotic figures or breast cancer non-mitotic figures.

\noindent \textbf{MIDOG Neuroendocrine.}
The MIDOG Neuroendocrine dataset \cite{aubreville2024domain} is a dataset for human pancreatic and gastrointestinal neuroendocrine tumors, characterized by their aggregated cellular morphology, obtained from the UMC Utrecht pathology archive. The dataset consisted of 55 WSI cases uniformly digitized using a Hamamatsu XR  scanner at 40$\times$ magnification.  We divide the slices into 1024$\times$1024 ROIs, resulting in 1,448 and 354 images for training and testing, respectively.

\noindent \textbf{Rings.}
The Rings dataset \cite{salvi2021hybrid} comprises WSIs of prostate biopsy specimens from 150 male patients, collected at the Division of Pathology, Department of Oncology (Turin, Italy). Tissue samples are formalin-fixed, paraffin-embedded, and sectioned, followed by hematoxylin and eosin (H\&E) staining. 
Two expert pathologists manually annotated gland contours, categorizing each into two classes: prostate healthy glands and prostate tumors. The dataset includes 43,739 annotated ROIs, split into 29,451 training and 14,288 test samples.

\noindent \textbf{SICAPv2.}
The SICAPv2 dataset \cite{silva2020going} comprises 155 prostate biopsy WSIs obtained from 95 consenting patients, acquired using a Ventana iScan Coreo scanner at 40$\times$ magnification. Expert urogenital pathologists at Hospital Clínico of Valencia performed comprehensive histological analyses of the specimens. The dataset contains 23,924 annotated ROIs, strategically partitioned into 19,140 training samples and 4,784 test samples to ensure robust model development and evaluation.

\noindent \textbf{TNBC.}
The TNBC dataset \cite{naylor2018segmentation} comprises 167 annotated H\&E-stained histopathology ROIs derived from triple-negative breast cancer specimens, systematically categorized into seven distinct cellular classes: glial, mitotic, cancerous, fibroblasts, lymphocyte plasmocyte, adipocytes, and endothelial. Captured at 40$\times$ magnification using a Philips Ultra Fast Scanner 1.6RA, the dataset spans eleven TNBC patients with 3-8 ROIs per case, deliberately sampling both tumor-rich regions and stromal microenvironments to ensure comprehensive representation of cellular diversity.
Annotations are performed by a team of three experts (including one board-certified pathologist) through a rigorous multi-stage protocol: initial marking in ITK-SNAP with zoom-assisted nuclear boundary delineation, cross-validation by a second annotator, and final consensus resolution for disputed cases. The dataset is partitioned into 134 training ROIs and 33 test ROIs.

\noindent \textbf{WSS4LUAD.}
The WSS4LUAD dataset \cite{han2022wsss4luad} consists of H\&E-stained WSIs of lung adenocarcinoma samples obtained from Guangdong Provincial People's Hospital (GDPH) and TCGA, designed for three-class tissue classification into tumor epithelial tissue, tumor-associated stroma tissue, and normal tissue. The training set contains 4,590 quality-controlled ROIs cropped from 49 GDPH and 14 TCGA WSIs. The test set comprises 1,145 ROIs from 9 GDPH and 3 TCGA WSIs.

\noindent \textbf{GlaS*.}
The GlaS dataset \cite{sirinukunwattana2017gland} consists of 165 histopathology images derived from 16 H\&E-stained WSIs of stage T3\/T4 colorectal adenocarcinoma, with each WSI originating from a distinct patient to ensure substantial inter-subject variability in both tissue architecture and staining characteristics. We employed 80 samples for external validation.

\noindent \textbf{NuCLS*.}
The NuCLS dataset \cite{amgad2022nucls} is a large-scale, multi-class histopathology dataset featuring expert-annotated nuclear boundaries across ten distinct nuclear types: lymphocyte, vascular endothelium, ductal epithelium, neutrophil, macrophage, tumor cells, apoptotic bodies, fibroblasts, plasma cells, and mitotic figures. Constructed through a structured crowdsourcing framework involving medical students and certified pathologists, the dataset originates from 125 triple-negative breast cancer (TNBC) WSIs. The dataset contains 905 samples for external validation.

\subsubsection*{ROI-Level VQA}

\textbf{PathMMU.} 
The PathMMU dataset \cite{sun2024pathmmu} represents a comprehensive multimodal resource for pathology VQA, aggregating data from five distinct sources: (1) PathMed (PubMed scientific articles), (2) Atlas (pathology textbooks), (3) EduContent (educational YouTube videos), (4) SocialPath (expert-contributed social media posts), and (5) PathCLS (existing pathology classification datasets). Due to access restrictions on SocialPath and training data, our study utilizes the official testtiny and validation subsets from the remaining four sources for evaluation, with 6,901 samples allocated for training.  Detailed experimental results are provided in Extended Data Table \ref{tab:ROI:vqa}.

\noindent \textbf{PathVQA.} 
The PathVQA \cite{he2020pathvqa} dataset provides 32,799 question-answer pairs (50.2\% open-ended, 49.8\% binary) associated with 4,998 pathological images from textbooks and the PEIR digital library. The dataset covers diverse query types (what, where, how, yes/no) to comprehensively assess model capabilities. We employ the official split of 19,654 training, 6,259 validation, and 6,719 test question-answer pairs for method development and evaluation. Detailed experimental results are provided in Extended Data Table \ref{tab:ROI:vqa}.

\subsubsection*{WSI-Level Image-Text Pairs}
\textbf{SlideInstruction-Caption.}
The SlideInstruction dataset \cite{chen2025slidechat} addresses the critical gap in large-scale multimodal resources for WSI understanding by providing 4,915 curated WSI-report pairs sourced from the TCGA database, spanning 4,028 patients. SlideInstruction-Caption Data, featuring concise and clinically focused summaries generated via GPT-4, highlights diagnostic findings while filtering noise (e.g., administrative text, specimen handling details).
\subsubsection*{WSI-Level Classification}

\noindent \textbf{TCGA.}
The Cancer Genome Atlas (TCGA) dataset \cite{weinstein2013cancer} serves as a comprehensive resource for cancer classification, integrating histopathological WSIs with expert annotations across multiple cancer types. For this study, we focus on six major malignancies: breast invasive carcinoma (BRCA), colon adenocarcinoma (COAD), head and neck squamous cell carcinoma (HNSC), lower-grade glioma (LGG), lung adenocarcinoma (LUAD), and lung squamous cell carcinoma (LUSC). Detailed experimental results are provided in Extended Data Table \ref{tab:WSI:tcga}.

\noindent \textbf{BRACS* (7 classes).}
We utilized the Breast Carcinoma Subtyping (BRACS) dataset \cite{brancati2022bracs} for external validation, which comprises 547 H\&E-stained WSIs of breast carcinoma specimens obtained from 187 patients. Following rigorous quality control to exclude slides with insufficient tumor proportion, a final cohort of 545 WSIs was retained for analysis. The BRACS dataset is designed for fine-grained classification of breast lesions, encompassing seven distinct histological subtypes: Normal, Pathological Benign, Usual Ductal Hyperplasia (UDH), Flat Epithelial Atypia (FEA), Atypical Ductal Hyperplasia (ADH), Ductal Carcinoma In Situ (DCIS), and Invasive Carcinoma. This dataset provides a comprehensive representation of the spectrum of breast lesions, ranging from benign and premalignant conditions to malignant neoplasms, making it suitable for evaluating the generalizability of computational pathology models. Detailed experimental results are provided in Extended Data Table \ref{tab:WSI:BRACS}.

\noindent \textbf{CAMELYON* (2 classes).}
CAMELYON dataset \cite{bejnordi2017diagnostic,bandi2018detection} comprises 899 whole-slide images (WSIs) of lymph node tissue stained with H\&E. This dataset is derived from two publicly available challenges: CAMELYON16 (399 slides) and CAMELYON17 (500 slides). The slides are categorized into two classes: normal (557 slides) and metastasis (341 slides). Following preprocessing, one corrupted normal slide was excluded, resulting in a final cohort of 898 WSIs for analysis. The CAMELYON dataset is widely recognized for benchmarking computational models in breast cancer metastasis detection, providing a robust external validation set to assess model generalizability.
Detailed experimental results are provided in Extended Data Table \ref{tab:WSI:CAMELYON}.

\noindent \textbf{CPTAC-NSCLC* (2 classes).}
For non-small cell lung cancer (NSCLC) subtyping, we employed data from the Clinical Proteomic Tumor Analysis Consortium (CPTAC) \cite{edwards2015cptac}, a comprehensive multi-omics resource integrating genomic, proteomic, and clinical data for cancer research 5. The CPTAC cohort includes 1,136 lung adenocarcinoma (LUAD) and 1,077 lung squamous cell carcinoma (LUSC) whole-slide images (WSIs), providing a robust dataset for binary classification tasks.
Detailed experimental results are provided in Extended Data Table \ref{tab:WSI:CPTAC-NSCLC}.

\subsubsection*{WSI-Level VQA}

\textbf{SlideInstruction-VQA.}
SlideInstruction-VQA comprises 175,753 VQA pairs structured into 3 broad classes (microscopy, diagnosis, clinical considerations) and 13 narrow subclasses (e.g., histopathological changes, differential diagnosis, and biomarker analysis ). The curation pipeline leverages GPT-4 to refine TCGA reports and generate diverse QA pairs while excluding low-relevance samples. To mimic real-world variability, the training set intentionally retains noisy cases where reports map ambiguously to multiple WSIs, enhancing model robustness. The test set strictly adheres to one-to-one WSI-report alignments for reliable evaluation. Detailed experimental results are provided in Extended Data Tables \ref{tab:slide_vqa:microscopy}-\ref{tab:slide_vqa:clinical}.

\noindent \textbf{BCNB*.}
The BCNB (Early Breast Cancer Core-Needle Biopsy WSI) dataset \cite{xu2021predicting} is a histopathology resource comprising 1,058 patients with early-stage breast cancer, featuring WSIs and corresponding clinical annotations verified by two independent pathologists. We utilize the dataset for external validation across four key VQA tasks: Grading (histological differentiation into Grades 1–3), HER2 Expression (positive/negative by IHC/ISH guidelines), HER2 Type (subclassification of HER2-low and HER2-positive cases), and Tumor Type (histological subtypes). Detailed experimental results are provided in Extended Data Table \ref{tab:slide_vqa:bcnb}.

\subsection{Compared Methods}

Our study compares the performance of our model with five open-sourced multimodal models: Qwen2.5-VL, a general purpose vision-language model; LLaVA-Med, a biomedical-specific model trained on a large corpus of medical literature; Quilt-LLaVA, a histopathology-focused model fine-tuned with spatially-grounded question-answer pairs; PathoR1, a contemporary work using reinforcement learning for reasoning refinement; and BiomedParse, a foundation model for joint segmentation and detection across various imaging modalities. By evaluating these models across a range of tasks, including ROI-level classification, detection, segmentation, VQA, WSI-level classification, and VQA, we aim to provide a comprehensive assessment of their strengths and weaknesses in computational pathology. This comparative analysis highlights the importance of domain-specific adaptations and the limitations of current methods in achieving robust generalizability across diverse histopathology applications.

\noindent \textbf{Qwen2.5-VL.}
Qwen2.5-VL \cite{bai2025qwen2} constitutes a general purpose vision-language model that advances multimodal understanding through several architectural innovations, including a dynamic resolution ViT encoder and optimized SwiGLU/RMSNorm components. The model demonstrates exceptional capability in document parsing and video temporal understanding, supported by its omnidocument processing framework that handles complex layouts, tables, and mathematical formulas. While not specifically optimized for medical applications, Qwen2.5-VL's robust multilingual support and object grounding capabilities make it suitable for preliminary medical image analysis tasks, particularly in resource-constrained settings.

\noindent \textbf{LLaVA-Med.}
LLaVA-Med \cite{li2023llava} emerges as a biomedical multimodal large language model specifically designed for medical image understanding and reasoning, developed through curriculum learning in the extensive collection of biomedical literature from PubMed Central. The model architecture builds upon LLaVA's foundation while incorporating domain-specific adaptations through two key phases: biomedical concept alignment using 1.6 million image-caption pairs from PMC-15M dataset, followed by instruction tuning with GPT-4 generated medical question-answer pairs. This approach enables LLaVA-Med to achieve state-of-the-art performance on biomedical VQA benchmarks.

\noindent \textbf{Quilt-LLaVA.}
Quilt-LLaMA \cite{quiltllava} represents a specialized vision-language model for histopathology analysis, built upon the LLaVA framework with significant domain-specific adaptations. The model employs a two-stage training approach: initial vision-text alignment using 107K curated histopathology question-answer pairs (Quilt-Instruct) derived from educational YouTube videos, followed by instruction tuning with spatially-grounded QA pairs generated through mouse cursor tracking.

\noindent \textbf{PathoR1.}
PathoR1 \cite{zhang2025patho}, a contemporary work, demonstrates a promising approach to improving reasoning in pathology VLMs through a three-stage pipeline involving pretraining, supervised fine-tuning with chain-of-thought examples, and reinforcement learning for reasoning refinement. However, the reinforcement learning is limited in scope, utilizing only 10K samples on ROI-level VQA tasks. It lacks capability to classification, detection, segmentation, or WSI-level tasks, potentially limiting its generalizability to these more complex tasks, a limitation that our current study addresses.

\noindent \textbf{BiomedParse.}
BiomedParse \cite{zhao2025foundation} establishes a new paradigm in medical image analysis as a comprehensive foundation model for joint segmentation and detection across nine imaging modalities. Its innovative architecture combines a FocalNet-based image encoder with PubMedBERT text understanding and a SEEM-inspired mask decoder, enabling text-prompted segmentation without requiring bounding box inputs. Trained on an unprecedented collection of over 6 million image-mask-text triples from 45 diverse medical segmentation datasets, the model demonstrates remarkable generalization capability.

\subsection{Evaluation Metrics}
We employ task-specific evaluation metrics to comprehensively assess model performance across different tasks.

\noindent \textbf{Classification Tasks.}
For classification tasks, we employ Accuracy (ACC) and F1-score as evaluation metrics. Accuracy measures the proportion of correctly classified samples, while the F1-score balances precision and recall.
\begin{equation}
\begin{split}
&\text{ACC} = \frac{1}{N} \sum_{i=1}^{N} \mathbb{I}(y_i = \hat{y}_i) = \frac{\text{TP} + \text{TN}}{\text{TP} + \text{TN} + \text{FP} + \text{FN}},\\
&\text{F1} = \frac{2 \times \text{Precision} \times \text{Recall}}{\text{Precision} + \text{Recall}} = \frac{2\text{TP}}{2\text{TP} + \text{FP} + \text{FN}},
\end{split}
\end{equation}
where $N$ is the total number of samples,$\mathbb{I}()$ is the indicator function, $y_i$ and $\hat{y}_i$ denote the ground truth and predicted labels, respectively, and TP/TN/FP/FN represent true positives, true negatives, false positives, and false negatives.

\noindent \textbf{Detection Tasks.}
Object detection performance is evaluated using Average Precision (AP) at different Intersection-over-Union (IoU) thresholds. We report AP30, AP50, and AP70 for specific threshold analysis.
\begin{equation}
\text{AP} = \int_0^1 p(r) dr,
\end{equation}
where $p(r)$ is the precision-recall curve.

\noindent \textbf{Segmentation Tasks.}
The Dice score, which measures the spatial overlap between predicted and ground truth masks, is used for segmentation evaluation:
\begin{equation}
\text{Dice} = \frac{2|X \cap Y|}{|X| + |Y|},
\end{equation}
where $X$ and $Y$ represent the predicted and ground truth masks, respectively.

\noindent \textbf{VQA Tasks.}
For close-ended VQA tasks, we use Accuracy (ACC). For open-ended tasks, we employ BLEU-4 to measure the n-gram overlap between generated and reference answers:
\begin{equation}
\text{BLEU-4} = BP \cdot \exp\left(\sum_{n=1}^4 w_n \log p_n\right),
\end{equation}
where $BP$ is the brevity penalty, $w_n$ are uniform weights, and $p_n$ are modified n-gram precisions.

\subsection{Implementation Details}
The model is implemented with PyTorch \cite{paszke2019pytorch} and trained on a 8$\times$80GB H800 GPU node. $M$, the token count per pathology image, is set to 256 for ROI-level tasks and 1024 for WSI-level tasks, respectively. The patch size $P$ is set to 28. The intrinsic rank and global scaling factor in LoRA are set to 32 and 128. AdamW \cite{kingma2014adam} is used as the optimizer with a weight decay of 0.1. The initial learning rate is set to 1e-4 for scale-dependent supervised fine-tuning and 1e-5 for task-aware reinforcement fine-tuning with a cosine learning rate schedule. KL regularization term coefficient $\beta$, clip coefficient $\epsilon$, and the number of sampled outputs $G$ in a group for GRPO are set to 0.001, 0.2, and 8, respectively.

\section*{Data Availability}
All datasets in the SmartPath-R1 are publicly available, with a full list of dataset links and statistical details being presented in Extended Data Tables \ref{tab:data_details_WSI}-\ref{tab:data_links}.

\section*{Code Availability}
The code for SmartPath-R1 will be released at \url{https://github.com/zhexu1997/SmartPath-R1}. For the competing methods, we adopt the official implementations of Qwen2.5VL (\url{https://github.com/QwenLM/Qwen2.5-VL}), LLaVA-Med (\url{https://github.com/microsoft/LLaVA-Med}), Quilt-LLaVA (\url{https://github.com/aldraus/quilt-llava}), PathoR1 (\url{https://github.com/wenchuan-zhang/patho-r1}), and BiomedParse (\url{https://github.com/microsoft/BiomedParse}).

\section*{Ethics Declarations}
This project has been reviewed and approved by the Human and Artefacts Research Ethics Committee (HAREC) of Hong Kong University of Science and Technology. The protocol number is HREP-2025-0163.

\section*{Author Contribution}
Z.X. and H.C. conceived and designed the work. Z.X., Z.L., J.H., J.M., C.J., Y.W., Z.C., Z.G., F.Z., and Y.X., curated the data included in the paper. Z.X. contributed to the technical implementation of the SmartPath-R1 framework and evaluated the performance of MLLMs.  Z.X. wrote the manuscript with inputs from all authors. All authors reviewed and approved the final paper. H.C. supervised the research.

\section*{Acknowledgment}
This work is supported by the National Natural Science Foundation of China (No. 62202403), Hong Kong Innovation and Technology Commission (Project No. MHP/002/22 and ITCPD/17-9), Shenzhen Science and Technology Innovation Committee Fund (Project No. KCXFZ20230731094059008) and Research Grants Council of the Hong Kong Special Administrative Region, China (Project No. R6003-22 and C4024-22GF).

\bibliography{sample.bib}

\section*{Extended Data}

\begin{table}[htbp]
\centering
\caption{\textbf{Classes of Different Datasets for ROI-Level Classification.} * represents external validation datasets.
}\label{tab:ROI:classes_cls}%
\begin{tabular}{lc}
\toprule
 \textbf{Dataset} & \textbf{Class}  \\
 \midrule
CCRCC &Renal cancer; Normal renal; \\
&Stromal, including smooth muscle, fibrous stroma and blood vessels; Red blood cells \\
Chaoyang & Normal mucosa; Serrated lesions; Adenocarcinoma; Adenom \\
CRC-MSI &MSIH, high microsatellite instability; \\
&nonMSIH, either low microsatellite instability or microsatel-lite stable\\
ESCA &Adventitia; Lamina propria mucosae; Muscularis mucosae; Muscularis propria; Regression tissue; \\ 
&Mucosa gastric; Mucosa oesophagus; Submucosa; Submucosal glands; Tumor; Ulceration\\
 PanCancerTIL  &TIL-negative, no significant lymphocyte infiltration;\\
 &TIL-positive, there are at least two TILs in the image\\
UniToPatho &Normal tissue; Hyperplastic polyp; \\
&Tubular adenoma, low-grade dysplasia; Tubular adenoma, high-grade dysplasia; \\
&Tubulovillous adenoma, low-grade dysplasia; Tubulovillous adenoma, high-grade dysplasia\\
BreakHis* & Benign tumor; Malignant tumor\\
 PanCancerTCGA* &Glioblastoma multiforme;  Kidney renal papillary cell carcinoma;\\ 
 & Colon adenocarcinoma; Lung squamous cell carcinoma; \\
 &Esophageal carcinoma; Uterine corpus endometrial carcinoma;\\
 &Bladder urothelial carcinoma; Brain lower grade glioma;\\ 
 &Cervical squamous cell carcinoma and endocervical adenocarcinoma; Ovarian serous cystadenocarcinoma; \\
 &Lymphoid neoplasm diffuse large b-cell lymphoma; Cholangiocarcinoma;\\
 &Liver hepatocellular carcinoma; Kidney renal clear cell carcinoma; \\
 &Skin cutaneous melanoma; Breast invasive carcinoma;\\
 &Stomach adenocarcinoma; Kidney chromophobe; Rectum adenocarcinoma; Mesothelioma;\\
 &Sarcoma; Lung adenocarcinoma; Pancreatic adenocarcinoma; Adrenocortical carcinoma;\\ 
 &Thymoma; Uterine carcinosarcoma; Pheochromocytoma and paraganglioma; Uveal melanoma;\\
 &Thyroid carcinoma; Testicular germ cell tumors; \\
 &Head and neck squamous cell carcinoma; Prostate adenocarcinoma\\
\bottomrule
\end{tabular}
\end{table}

\begin{table}[htbp]
\centering
\caption{\textbf{Average ROI-Level Classification Performance of MLLMs across 8 Tasks.}
Best performing model for each metric is \textbf{bolded} and second-best performing model is \underline{underlined}. The 95\% CI is included in parentheses.
}\label{tab:ROI:avg_cls}%
\begin{tabular}{lc}
\toprule
 \textbf{Method} & \textbf{ACC}  \\
 \midrule
Qwen2.VL&0.424 (0.235, 0.624) \\
LLaVA-Med&0.235 (0.144, 0.326)\\
Quilt-LLaVA&0.290 (0.149, 0.431)\\
PathoR1&0.275 (0.105, 0.445)\\
SmartPath-SFT &\underline{0.766 (0.644, 0.889)}\\
SmartPath-R1&\textbf{0.806 (0.706, 0.906)}\\
\bottomrule
\end{tabular}
\end{table}

\begin{table}[htbp]
\centering
\caption{\textbf{ROI-Level Classification Performance on CCRCC Dataset.} 
Best performing model is \textbf{bolded} and second-best is \underline{underlined}. The 95\% CI is included in parentheses.}
\label{tab:ROI:CCRCC}
\begin{tabular}{lccc}
\toprule
\textbf{Method} &\textbf{\# Class} & \textbf{ACC} & \textbf{F1}  \\
\midrule
Qwen2.VL &4  & 0.412 (0.391, 0.434)&0.392 (0.347,0.406) \\
LLaVA-Med &4   & 0.333 (0.313, 0.354)& 0.218 (0.195,0.229) \\
Quilt-LLaVA &4    & 0.384 (0.363, 0.406) & 0.270 (0.263,0.298)\\
PathoR1 &4    & 0.422 (0.400, 0.444) & 0.201 (0.201,0.235)\\
SmartPath-SFT &4   & \underline{0.907 (0.895, 0.920)}& \underline{0.888 (0.863,0.903)}\\
SmartPath-R1 &4   & \textbf{0.928 (0.917, 0.939)} &\textbf{0.895 (0.872,0.919)}\\
\bottomrule
\end{tabular}
\end{table}

\begin{table}[htbp]
\centering
\caption{\textbf{ROI-Level Classification Performance on Chaoyang Dataset.} 
Best performing model is \textbf{bolded} and second-best is \underline{underlined}. The 95\% CI is included in parentheses.}
\label{tab:ROI:Chaoyang}
\begin{tabular}{lccc}
\toprule
\textbf{Method} &\textbf{\# Class} & \textbf{ACC} & \textbf{F1}\\
\midrule
  Qwen2.VL &4   & 0.438 (0.407, 0.469) &0.325 (0.314,0.368)\\
  LLaVA-Med &4   & 0.316 (0.287, 0.345) &0.254 (0.225,0.279))\\
  Quilt-LLaVA &4   & 0.219 (0.193, 0.245) & 0.206 (0.204,0.255)\\
PathoR1 &4    &   0.300 (0.272, 0.328) & 0.242 (0.200,0.252)\\
  SmartPath-SFT &4  & \underline{0.765 (0.739, 0.791)} & \underline{0.662 (0.644,0.711)}\\
  SmartPath-R1 &4  & \textbf{0.795 (0.770, 0.820)} &\textbf{0.724 (0.676,0.738)} \\
\bottomrule
\end{tabular}
\end{table}

\begin{table}[htbp]
\centering
\caption{\textbf{ROI-Level Classification Performance on CRCMSI Dataset.} 
Best performing model is \textbf{bolded} and second-best is \underline{underlined}. The 95\% CI is included in parentheses.}
\label{tab:ROI:CRCMSI}
\begin{tabular}{lccc}
\toprule
\textbf{Method} &\textbf{\# Class}  & \textbf{ACC} & \textbf{F1}\\
\midrule
 Qwen2.VL &2  & \underline{0.640 (0.619, 0.661)} &0.488 (0.471,0.514)\\
 LLaVA-Med &2  & 0.188 (0.170, 0.205) &0.478 (0.429,0.471)
\\
 Quilt-LLaVA &2  &0.430 (0.408, 0.452) &0.410 (0.397,0.437)\\
 PathoR1 &2    & 0.370 (0.349, 0.392) &0.418 (0.408,0.448)\\
 SmartPath-SFT  &2  & 0.572 (0.550, 0.593) &\underline{0.517 (0.501,0.543)}\\
 SmartPath-R1 &2  & \textbf{0.744 (0.725, 0.763)} &\textbf{0.634 (0.591,0.643)}\\
\bottomrule
\end{tabular}
\end{table}

\begin{table}[htbp]
\centering
\caption{\textbf{ROI-Level Classification Performance on ESCA Dataset.} 
Best performing model is \textbf{bolded} and second-best is \underline{underlined}. The 95\% CI is included in parentheses.}
\label{tab:ROI:ESCA}
\begin{tabular}{lccc}
\toprule
\textbf{Method} &\textbf{\# Class}  & \textbf{ACC}& \textbf{F1} \\
\midrule
  Qwen2.VL &11  & 0.284 (0.265, 0.304) &0.122 (0.108,0.129)\\
  LLaVA-Med &11 &0.201 (0.183, 0.219) & 0.075 (0.065,0.085)\\
  Quilt-LLaVA &11 &0.285 (0.266, 0.305) & 0.121 (0.096,0.128)\\
  PathoR1 &11    &   0.111 (0.097, 0.125) &0.079 (0.065,0.086) \\
  SmartPath-SFT &11 & \underline{0.883 (0.869, 0.897)} &\underline{0.608 (0.562,0.660)}\\
  SmartPath-R1 &11 & \textbf{0.895 (0.882, 0.908)} & \textbf{0.625 (0.586,0.719)}\\
\bottomrule
\end{tabular}
\end{table}

\begin{table}[htbp]
\centering
\caption{\textbf{ROI-Level Classification Performance on PanCancerTIL Dataset.} 
Best performing model is \textbf{bolded} and second-best is \underline{underlined}. The 95\% CI is included in parentheses.}
\label{tab:ROI:PanCancerTIL}
\begin{tabular}{lccc}
\toprule
\textbf{Method} &\textbf{\# Class}  & \textbf{ACC} & \textbf{F1}\\
\midrule
  Qwen2.VL &2   &  0.748 (0.728, 0.767) &0.494 (0.474,0.517)\\
  LLaVA-Med &2  & 0.266 (0.247, 0.285)& 0.445 (0.426,0.470)\\
  Quilt-LLaVA &2  &  0.361 (0.340, 0.383) & 0.449 (0.415,0.456)\\
  PathoR1 &2    & 0.196 (0.178, 0.213) &0.447 (0.420,0.462) \\
 SmartPath-SFT &2  & \underline{0.905 (0.893, 0.918)} &\underline{0.851 (0.835,0.873)}\\
  SmartPath-R1 &2  & \textbf{0.915 (0.903, 0.928)} &\textbf{0.889 (0.849,0.888)}\\
\bottomrule
\end{tabular}
\end{table}

\begin{table}[htbp]
\centering
\caption{\textbf{ROI-Level Classification Performance on UniToPatho Dataset.} 
Best performing model is \textbf{bolded} and second-best is \underline{underlined}. The 95\% CI is included in parentheses.}
\label{tab:ROI:UniToPatho}
\begin{tabular}{lccc}
\toprule
\textbf{Method} &\textbf{\# Class}  & \textbf{ACC} & \textbf{F1}\\
  \midrule
  Qwen2.VL &6  &  0.164 (0.141, 0.187) &0.137 (0.116,0.157)\\
  LLaVA-Med &6 &0.126 (0.105, 0.147) & 0.131 (0.112,0.153)\\
  Quilt-LLaVA &6 & 0.078 (0.061, 0.095) & 0.161 (0.143,0.190)\\
  PathoR1 &6    & 0.082 (0.065, 0.099) &0.148 (0.127,0.172) \\
 SmartPath-SFT &6 & \underline{0.541 (0.510, 0.572)} &\underline{0.433 (0.403,0.470)}\\
  SmartPath-R1 &6 & \textbf{0.580 (0.549, 0.611)} &\textbf{0.461 (0.420,0.485)}\\
\bottomrule
\end{tabular}
\end{table}

\begin{table}[htbp]
\centering
\caption{\textbf{External Validation of ROI-Level Classification Performance on BreakHis Dataset.} 
Best performing model is \textbf{bolded} and second-best is \underline{underlined}. The 95\% CI is included in parentheses.}
\label{tab:ROI:BreakHis}
\begin{tabular}{lccc}
\toprule
\textbf{Method} &\textbf{\# Class}  & \textbf{ACC} & \textbf{F1}\\
  \midrule
  Qwen2.VL &2  & 0.628 (0.598, 0.658) &0.492 (0.460,0.520)\\
  LLaVA-Med &2 & 0.384 (0.354, 0.414) & 0.480 (0.444,0.508)\\
  Quilt-LLaVA &2 & 0.522 (0.491, 0.553) & 0.513 (0.464,0.523)\\
  PathoR1 &2    &0.654 (0.624, 0.684) &0.493 (0.467,0.527)\\
 SmartPath-SFT &2 & \underline{0.842 (0.819, 0.865)} &\textbf{0.819 (0.802,0.851)}\\
  SmartPath-R1 &2 & \textbf{0.868 (0.847, 0.889)} &\underline{0.814 (0.821,0.869)}\\
\bottomrule
\end{tabular}
\end{table}

\begin{table}[htbp]
\centering
\caption{\textbf{External Validation of ROI-Level Classification Performance on PanCancerTCGA Dataset.} 
Best performing model is \textbf{bolded} and second-best is \underline{underlined}. The 95\% CI is included in parentheses.}
\label{tab:ROI:PanCancerTCGA}
\begin{tabular}{lccc}
\toprule
\textbf{Method} &\textbf{\# Class}  & \textbf{ACC} & \textbf{F1}\\
  \midrule
  Qwen2.VL &32  & 0.073 (0.062, 0.085) &0.029 (0.025,0.036)\\
  LLaVA-Med &32 &0.066 (0.055, 0.077) & 0.026 (0.025,0.039)\\
  Quilt-LLaVA &32 &0.042 (0.033, 0.051)& 0.022 (0.018,0.031)\\
  PathoR1 &32    & 0.065 (0.054, 0.075) &0.029 (0.028,0.043) \\
 SmartPath-SFT &32 & \underline{0.715 (0.695, 0.734)} &\underline{0.613 (0.590,0.652)}\\
  SmartPath-R1 &32 & \textbf{0.722 (0.702, 0.742)} &\textbf{0.627 (0.607,0.662)}\\
\bottomrule
\end{tabular}
\end{table}

\begin{table}[htbp]
\centering
\caption{\textbf{Classes of Different Datasets for ROI-Level Detection and Segmentation.} * represents external validation datasets.
}\label{tab:ROI:classes_det}%
\begin{tabular}{lc}
\toprule
 \textbf{Dataset} & \textbf{Class}  \\
 \midrule
BCSS &Mucoid material; Blood; Metaplasia NOS; Glandular secretions; \\
&Necrosis or debris; Plasma cells; Lymphatics; Blood vessel; \\
&Other immune; Dcis; Tumor; Stroma; \\
&Normal acinus or duct; Fat; Lymphocytic; Angioinvasion \\
CoCaHis &Metastatic colon cancer\\
CoNIC2022 &Neutrophil; Connective; Lymphocyte; Plasma; Eosinophil; Epithelial\\
CRAG &Gland \\
DigestPath2019-Tissue & Malignant lesion \\
DigestPath2019-Cell &Signet ring cell\\
MIDOG-Breast &Breast cancer non mitotic figure; Breast cancer mitotic figure\\
MIDOG-Neuroendocrine &Neuroendocrine tumor mitotic figure; Neuroendocrine tumor non mitotic figure\\
Rings &prostate healthy glands; Prostate tumor\\
SICAPv2 &Prostate cancer \\
TNBC &Lymphocyte plasmocyte; Mitosis; Fibroblast; Cancerous; Adipocyte; Glial; Endothelial\\
WSSS4LUAD &Normal tissue; Tumor associated stroma tissue; Tumor epithelial tissue \\
GlaS* &Gland\\
NuCLS* &Macrophage; Neutrophil; Apoptotic body; Fibroblast; Lymphocyte;  \\
& Vascular endothelium; Ductal epithelium; Tumor; Mitotic figure; Plasma cell\\
\bottomrule
\end{tabular}
\end{table}

\begin{figure*}
    \centering
    \includegraphics[width=1\linewidth]{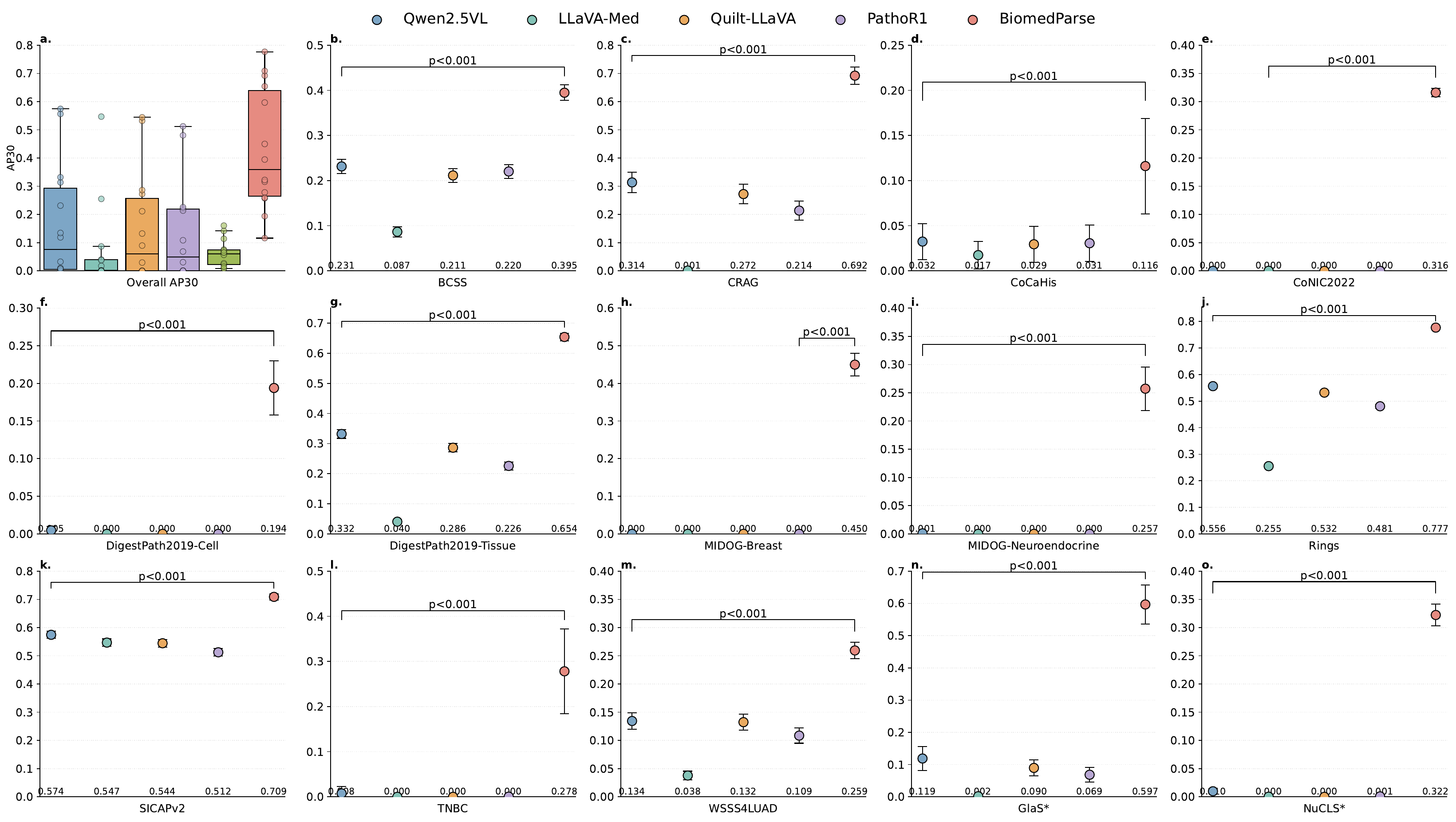}
\caption{\textbf{Performance of MLLMs on ROI-Level Detection Tasks.}
\textbf{a.} Average performance of MLLMs based on average precision at IoU threshold 0.3 (AP30) across 14 ROI-level detection tasks.
\textbf{b-o.} Model performance on specific tasks. * represents external validation datasets. Error bars represent 95\% CI. The box limits represent the standard error. P-values are computed using a Wilcoxon signed-rank two-sided test. 
}
\label{fig:ROI_det_30}
\end{figure*}

\begin{figure*}
    \centering
    \includegraphics[width=1\linewidth]{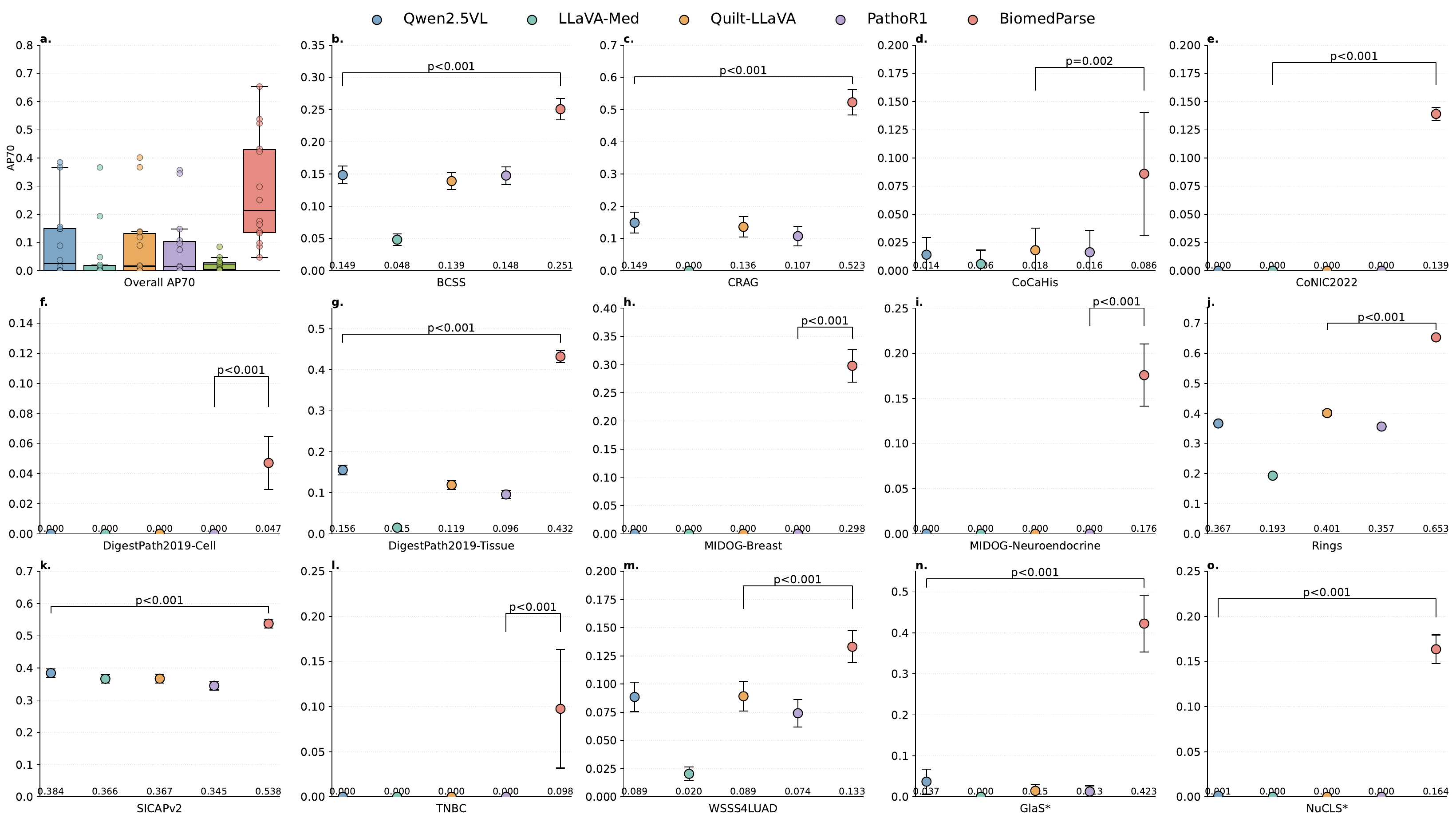}
\caption{\textbf{Performance of MLLMs on ROI-Level Detection Tasks.}
\textbf{a.} Average performance of MLLMs based on average precision at IoU threshold 0.7 (AP70) across 14 ROI-level detection tasks.
\textbf{b-o.} Model performance on specific tasks. * represents external validation datasets. Error bars represent 95\% CI. The box limits represent the standard error. P-values are computed using a Wilcoxon signed-rank two-sided test. 
}
\label{fig:ROI_det_70}
\end{figure*}

\begin{table}[htbp]
\centering
\caption{\textbf{Average ROI-Level Detection Performance on 14 datasets.} 
Best performing model is \textbf{bolded} and second-best is \underline{underlined}. The 95\% CI is included in parentheses.}
\label{tab:ROI_det:avg}
\begin{tabular}{lccc}
\toprule
\textbf{Method}  & \textbf{AP30} & \textbf{AP50} & \textbf{AP70}\\
  \midrule
  Qwen2.VL  &0.165 (0.047, 0.284)  &0.127 (0.030, 0.225) & 0.096 (0.019, 0.173)  \\
  LLaVA-Med &0.071 (0.000, 0.159) &0.057 (0.000, 0.132) &0.046 (0.000, 0.107)\\
  Quilt-LLaVA  & 0.150 (0.038, 0.262) & 0.117 (0.023, 0.212) &0.092 (0.014, 0.170)\\
PathoR1  &0.133 (0.031, 0.235) & 0.104 (0.018, 0.191)&0.083 (0.011, 0.154)\\
  BiomedParse &0.061 (0.033, 0.089) &0.039 (0.020, 0.059) &0.023 (0.010, 0.036) \\
 SmartPath-SFT  &\underline{0.352 (0.228, 0.475)} &\underline{0.312 (0.195, 0.428)} &\underline{0.233 (0.217, 0.249)}\\
  SmartPath-R1  &\textbf{0.430 (0.305, 0.555)} &\textbf{0.376 (0.257, 0.494)} & \textbf{0.283 (0.170, 0.396)}\\
\bottomrule
\end{tabular}
\end{table}

\begin{table}[htbp]
\centering
\caption{\textbf{ROI-Level Detection Performance on BCSS Dataset.} 
Best performing model is \textbf{bolded} and second-best is \underline{underlined}. The 95\% CI is included in parentheses.}
\label{tab:ROI_det:BCSS}
\begin{tabular}{lccc}
\toprule
\textbf{Method}  & \textbf{AP30} & \textbf{AP50} & \textbf{AP70}\\
  \midrule
  Qwen2.VL  &0.231 (0.216, 0.247) &0.186 (0.171, 0.201) &0.149 (0.135, 0.162)  \\
  LLaVA-Med &0.087 (0.075, 0.098) &0.062 (0.052, 0.071) &0.048 (0.040, 0.057)\\
  Quilt-LLaVA  &0.211 (0.196, 0.227) &0.169 (0.154, 0.183) &0.139 (0.126, 0.152)\\
PathoR1  &0.220 (0.205, 0.236) & 0.176 (0.161, 0.190) &0.148 (0.134, 0.161) \\
  BiomedParse &0.023 (0.020, 0.026) &0.016 (0.013, 0.019)&0.010 (0.007, 0.012) \\
 SmartPath-SFT  &\underline{0.339 (0.322, 0.357)} &\underline{0.292 (0.275, 0.309)} &\underline{0.233 (0.217, 0.249)}\\
  SmartPath-R1  &\textbf{0.395 (0.377, 0.412)} &\textbf{0.327 (0.310, 0.345)} & \textbf{0.251 (0.234, 0.267)}\\
\bottomrule
\end{tabular}
\end{table}

\begin{table}[htbp]
\centering
\caption{\textbf{ROI-Level Detection Performance on CRAG Dataset.} 
Best performing model is \textbf{bolded} and second-best is \underline{underlined}. The 95\% CI is included in parentheses.}
\label{tab:ROI_det:CRAG}
\begin{tabular}{lccc}
\toprule
\textbf{Method} & \textbf{AP30} & \textbf{AP50} & \textbf{AP70} \\
  \midrule
  Qwen2.VL &0.314 (0.278, 0.349) &0.228 (0.193, 0.263) &0.149 (0.116, 0.181)\\
  LLaVA-Med &0.001 (0.000, 0.003) &0.000 (0.000, 0.000) &0.000 (0.000, 0.000)\\
  Quilt-LLaVA  &0.272 (0.238, 0.307) &0.196 (0.162, 0.230) &0.136 (0.104, 0.168)\\  
  PathoR1  & 0.214 (0.180, 0.247) & 0.152 (0.120, 0.184) &0.107 (0.077, 0.137) \\
  BiomedParse &0.066 (0.057, 0.075) &0.046 (0.038, 0.054) &0.029 (0.024, 0.034) \\
 SmartPath-SFT  &\underline{0.592 (0.556, 0.627)} &\underline{0.542 (0.503, 0.581)} &\underline{0.459 (0.417, 0.501)}\\
  SmartPath-R1  &\textbf{0.692 (0.661, 0.723)} &\textbf{0.620 (0.585, 0.656)} &\textbf{0.523 (0.484, 0.562)}\\
\bottomrule
\end{tabular}
\end{table}

\begin{table}[htbp]
\centering
\caption{\textbf{ROI-Level Detection Performance on CoCaHis Dataset.} 
Best performing model is \textbf{bolded} and second-best is \underline{underlined}. The 95\% CI is included in parentheses.}
\label{tab:ROI_det:CoCaHis}
\begin{tabular}{lccc}
\toprule
\textbf{Method} & \textbf{AP30} & \textbf{AP50} & \textbf{AP70} \\
  \midrule
  Qwen2.VL & 0.032 (0.013, 0.052) &0.015 (0.000, 0.030) &0.014 (0.000, 0.029)\\
  LLaVA-Med &0.017 (0.002, 0.033) &0.006 (0.000, 0.018) &0.006 (0.000, 0.018)\\
  Quilt-LLaVA  &0.029 (0.010, 0.049) &0.019 (0.000, 0.039) &0.018 (0.000, 0.038)\\  
  PathoR1  & 0.031 (0.011, 0.051) & 0.017 (0.000, 0.037) &  0.016 (0.000, 0.036)\\
  BiomedParse &0.022 (0.010, 0.033) &0.017 (0.006, 0.028) &0.007 (0.000, 0.013) \\
 SmartPath-SFT  &\underline{0.097 (0.047, 0.147)} &\underline{0.074 (0.023, 0.124)} &\underline{0.064 (0.012, 0.115)}\\
  SmartPath-R1  &\textbf{0.116 (0.063, 0.169)} &\textbf{0.100 (0.046, 0.155)} &\textbf{0.086 (0.032, 0.140)}\\
\bottomrule
\end{tabular}
\end{table}

\begin{table}[htbp]
\centering
\caption{\textbf{ROI-Level Detection Performance on CoNIC2022 Dataset.} 
Best performing model is \textbf{bolded} and second-best is \underline{underlined}. The 95\% CI is included in parentheses.}
\label{tab:ROI_det:CoNIC2022}
\begin{tabular}{lccc}
\toprule
\textbf{Method} & \textbf{AP30} & \textbf{AP50} & \textbf{AP70} \\
  \midrule
  Qwen2.VL &0.000 (0.000, 0.000) &0.000 (0.000, 0.000) &0.000 (0.000, 0.000)\\
  LLaVA-Med &0.000 (0.000, 0.000) &0.000 (0.000, 0.000) &0.000 (0.000, 0.000)\\
  Quilt-LLaVA  &0.000 (0.000, 0.000) &0.000 (0.000, 0.000) &0.000 (0.000, 0.000)\\  
  PathoR1  &0.000 (0.000, 0.000)&0.000 (0.000, 0.000)&0.000 (0.000, 0.000) \\
  BiomedParse &0.076 (0.071, 0.080) &0.053 (0.049, 0.057) &0.027 (0.024, 0.030) \\
 SmartPath-SFT  &\underline{0.247 (0.240, 0.254)} &\underline{0.197 (0.190, 0.203)} &\underline{0.114 (0.108, 0.119)}\\
  SmartPath-R1  &\textbf{0.316 (0.309, 0.323)} &\textbf{0.252 (0.245, 0.259)} &\textbf{0.139 (0.133, 0.145)}\\
\bottomrule
\end{tabular}
\end{table}

\begin{table}[htbp]
\centering
\caption{\textbf{ROI-Level Detection Performance on DigestPath2019-Cell Dataset.} 
Best performing model is \textbf{bolded} and second-best is \underline{underlined}. The 95\% CI is included in parentheses.}
\label{tab:ROI_det:DigestPath2019-Cell}
\begin{tabular}{lccc}
\toprule
\textbf{Method} & \textbf{AP30} & \textbf{AP50} & \textbf{AP70} \\
  \midrule
  Qwen2.VL &0.000 (0.000, 0.000) &0.001 (0.000, 0.003) &0.000 (0.000, 0.000)\\
  LLaVA-Med &0.000 (0.000, 0.000) &0.000 (0.000, 0.000) &0.000 (0.000, 0.000)\\
  Quilt-LLaVA  &0.000 (0.000, 0.000) &0.000 (0.000, 0.000) &0.000 (0.000, 0.000)\\   
  PathoR1  &0.000 (0.000, 0.000) &0.000 (0.000, 0.000) &0.000 (0.000, 0.000) \\
  BiomedParse &0.000 (0.000, 0.000) &0.014 (0.009, 0.018) &0.004 (0.002, 0.007) \\
 SmartPath-SFT  &\underline{0.126 (0.096, 0.156)} &\underline{0.087 (0.064, 0.110)} &\underline{0.042 (0.027, 0.057)}\\
  SmartPath-R1  &\textbf{0.194 (0.158, 0.230)} &\textbf{0.122 (0.093, 0.151)} &\textbf{0.047 (0.030, 0.065)}\\
\bottomrule
\end{tabular}
\end{table}

\begin{table}[htbp]
\centering
\caption{\textbf{ROI-Level Detection Performance on DigestPath2019-Tissue Dataset.} 
Best performing model is \textbf{bolded} and second-best is \underline{underlined}. The 95\% CI is included in parentheses.}
\label{tab:ROI_det:DigestPath2019-Tissue}
\begin{tabular}{lccc}
\toprule
\textbf{Method} & \textbf{AP30} & \textbf{AP50} & \textbf{AP70} \\
  \midrule
  Qwen2.VL &0.332 (0.318, 0.346) &0.236 (0.222, 0.249) &0.156 (0.143, 0.168)\\
  LLaVA-Med &0.040 (0.034, 0.047) &0.022 (0.017, 0.027) &0.015 (0.011, 0.019)\\
  Quilt-LLaVA  &0.286 (0.272, 0.300) &0.191 (0.178, 0.204) &0.119 (0.108, 0.130)\\  
  PathoR1  & 0.226 (0.213, 0.239) &0.149 (0.138, 0.161) &0.096 (0.086, 0.106) \\
  BiomedParse &0.142 (0.134, 0.150) &0.115 (0.108, 0.123) &0.085 (0.078, 0.092) \\
 SmartPath-SFT  & \underline{0.541 (0.527, 0.554)} &\underline{0.469 (0.455, 0.484)} &\underline{0.374 (0.359, 0.388)}\\
  SmartPath-R1  &\textbf{0.654 (0.642, 0.667)} &\textbf{0.555 (0.541, 0.569)} &\textbf{0.432 (0.417, 0.447)}\\
\bottomrule
\end{tabular}
\end{table}

\begin{table}[htbp]
\centering
\caption{\textbf{ROI-Level Detection Performance on MIDOG-Breast Dataset.} 
Best performing model is \textbf{bolded} and second-best is \underline{underlined}. The 95\% CI is included in parentheses.}
\label{tab:ROI_det:MIDOG-Breast}
\begin{tabular}{lccc}
\toprule
\textbf{Method} & \textbf{AP30} & \textbf{AP50} & \textbf{AP70} \\
  \midrule
  Qwen2.VL &0.000 (0.000, 0.000) &0.000 (0.000, 0.000) &0.000 (0.000, 0.000)\\
  LLaVA-Med &0.000 (0.000, 0.000) &0.000 (0.000, 0.000) &0.000 (0.000, 0.000)\\
  Quilt-LLaVA  &0.000 (0.000, 0.000) &0.000 (0.000, 0.000) &0.000 (0.000, 0.000)\\  
  PathoR1  & 0.000 (0.000, 0.000) & 0.000 (0.000, 0.000) & 0.000 (0.000, 0.000) \\
  BiomedParse &0.009 (0.006, 0.012) &0.004 (0.002, 0.006) & 0.001 (0.000, 0.001)\\
 SmartPath-SFT  &\underline{0.367 (0.336, 0.398)} &\underline{0.348 (0.318, 0.379)} &\underline{0.256 (0.227, 0.284)}\\
  SmartPath-R1  &\textbf{0.450 (0.419, 0.480)} &\textbf{0.428 (0.398, 0.459)} &\textbf{0.298 (0.269, 0.327)}\\
\bottomrule
\end{tabular}
\end{table}

\begin{table}[htbp]
\centering
\caption{\textbf{ROI-Level Detection Performance on MIDOG-Neuroendocrine Dataset.} 
Best performing model is \textbf{bolded} and second-best is \underline{underlined}. The 95\% CI is included in parentheses.}
\label{tab:ROI_det:MIDOG-Neuroendocrine}
\begin{tabular}{lccc}
\toprule
\textbf{Method} & \textbf{AP30} & \textbf{AP50} & \textbf{AP70} \\
  \midrule
  Qwen2.VL &0.000 (0.000, 0.000) &0.000 (0.000, 0.000) &0.000 (0.000, 0.000)\\
  LLaVA-Med &0.000 (0.000, 0.000)&0.000 (0.000, 0.000) &0.000 (0.000, 0.000)\\
  Quilt-LLaVA  &0.000 (0.000, 0.000)&0.000 (0.000, 0.000) &0.000 (0.000, 0.000)\\   
  PathoR1  & 0.000 (0.000, 0.000) & 0.000 (0.000, 0.000) & 0.000 (0.000, 0.000) \\
  BiomedParse &0.008 (0.004, 0.013) &0.002 (0.000, 0.003) &0.001 (0.000, 0.001) \\
 SmartPath-SFT  &\underline{0.190 (0.153, 0.227)} &\underline{0.183 (0.147, 0.219)} &\underline{0.138 (0.105, 0.171)}\\
  SmartPath-R1  &\textbf{0.257 (0.219, 0.296)}&\textbf{0.236 (0.198, 0.273)} &\textbf{0.176 (0.141, 0.210)}\\
\bottomrule
\end{tabular}
\end{table}

\begin{table}[htbp]
\centering
\caption{\textbf{ROI-Level Detection Performance on Rings Dataset.} 
Best performing model is \textbf{bolded} and second-best is \underline{underlined}. The 95\% CI is included in parentheses.}
\label{tab:ROI_det:Rings}
\begin{tabular}{lccc}
\toprule
\textbf{Method} & \textbf{AP30} & \textbf{AP50} & \textbf{AP70} \\
  \midrule
  Qwen2.VL &0.556 (0.549, 0.564) &0.452 (0.444, 0.459) &0.367 (0.359, 0.374)\\
  LLaVA-Med & 0.255 (0.248, 0.262) &0.224 (0.217, 0.230) &0.193 (0.187, 0.199)\\
  Quilt-LLaVA  &0.532 (0.525, 0.539) &0.464 (0.457, 0.472) &0.401 (0.394, 0.409)\\   
  PathoR1  & 0.481 (0.473, 0.488) &0.418 (0.410, 0.425) &0.357 (0.349, 0.364) \\
  BiomedParse &0.072 (0.070, 0.074) & 0.047 (0.045, 0.048) & 0.028 (0.027, 0.029) \\
 SmartPath-SFT  &\underline{0.739 (0.732, 0.745)} &\underline{0.698 (0.692, 0.705)} &\underline{0.631 (0.624, 0.638)}\\
  SmartPath-R1  &\textbf{0.777 (0.771, 0.783)} &\textbf{0.732 (0.725, 0.738)} &\textbf{0.653 (0.646, 0.660)}\\
\bottomrule
\end{tabular}
\end{table}

\begin{table}[htbp]
\centering
\caption{\textbf{ROI-Level Detection Performance on SICAPv2 Dataset.} 
Best performing model is \textbf{bolded} and second-best is \underline{underlined}. The 95\% CI is included in parentheses.}
\label{tab:ROI_det:SICAPv2}
\begin{tabular}{lccc}
\toprule
\textbf{Method} & \textbf{AP30} & \textbf{AP50} & \textbf{AP70} \\
  \midrule
  Qwen2.VL &0.574 (0.562, 0.587) & 0.485 (0.472, 0.498) &0.384 (0.371, 0.398)\\
  LLaVA-Med &0.547 (0.534, 0.560) &0.456 (0.443, 0.469) &0.366 (0.353, 0.379)\\
  Quilt-LLaVA  &0.544 (0.531, 0.557) &0.455 (0.442, 0.469) &0.367 (0.354, 0.380)\\  
  PathoR1  &0.512 (0.499, 0.526) &0.428 (0.415, 0.441) & 0.345 (0.332, 0.358) \\
  BiomedParse &0.065 (0.061, 0.068) &0.042 (0.039, 0.045) &0.025 (0.023, 0.027) \\
 SmartPath-SFT  &\underline{0.686 (0.674, 0.697)} &\underline{0.613 (0.600, 0.625)} &\underline{0.521 (0.508, 0.535)}\\
  SmartPath-R1  &\textbf{0.709 (0.698, 0.720)} &\textbf{0.636 (0.624, 0.648)} &\textbf{0.538 (0.524, 0.551)}\\
\bottomrule
\end{tabular}
\end{table}

\begin{table}[htbp]
\centering
\caption{\textbf{ROI-Level Detection Performance on TNBC Dataset.} 
Best performing model is \textbf{bolded} and second-best is \underline{underlined}. The 95\% CI is included in parentheses.}
\label{tab:ROI_det:TNBC}
\begin{tabular}{lccc}
\toprule
\textbf{Method} & \textbf{AP30} & \textbf{AP50} & \textbf{AP70} \\
  \midrule
  Qwen2.VL &0.000 (0.000, 0.000) &0.000 (0.000, 0.000) &0.000 (0.000, 0.000)\\
  LLaVA-Med &0.000 (0.000, 0.000) &0.000 (0.000, 0.000) &0.000 (0.000, 0.000)\\
  Quilt-LLaVA  &0.000 (0.000, 0.000) &0.000 (0.000, 0.000) &0.000 (0.000, 0.000)\\  
  PathoR1  &0.000 (0.000, 0.000) &0.000 (0.000, 0.000) &0.000 (0.000, 0.000) \\
  BiomedParse &0.161 (0.088, 0.233) &0.078 (0.031, 0.125) &0.033 (0.006, 0.060) \\
 SmartPath-SFT  &\underline{0.173 (0.095, 0.250)} &\underline{0.162 (0.086, 0.238)} &\underline{0.112 (0.039, 0.184)}\\
  SmartPath-R1  &\textbf{0.278 (0.184, 0.372)} & \textbf{0.255 (0.163, 0.348)} & \textbf{0.098 (0.032, 0.163)}\\
\bottomrule
\end{tabular}
\end{table}

\begin{table}[htbp]
\centering
\caption{\textbf{ROI-Level Detection Performance on WSSS4LUAD Dataset.} 
Best performing model is \textbf{bolded} and second-best is \underline{underlined}. The 95\% CI is included in parentheses.}
\label{tab:ROI_det:WSSS4LUAD}
\begin{tabular}{lccc}
\toprule
\textbf{Method} & \textbf{AP30} & \textbf{AP50} & \textbf{AP70} \\
  \midrule
  Qwen2.VL &0.134 (0.120, 0.149) &0.110 (0.096, 0.124) &0.089 (0.075, 0.102)\\
  LLaVA-Med &0.038 (0.030, 0.045) &0.026 (0.019, 0.033) &0.020 (0.014, 0.026)\\
  Quilt-LLaVA  &0.132 (0.118, 0.147) &0.110 (0.096, 0.124) &0.089 (0.076, 0.102)\\   
  PathoR1  & 0.109 (0.095, 0.122) & 0.091 (0.078, 0.104) & 0.074 (0.062, 0.086)\\
  BiomedParse &0.015 (0.013, 0.017) &0.006 (0.005, 0.007) & 0.003 (0.002, 0.004) \\
 SmartPath-SFT  &\underline{0.209 (0.194, 0.223)} &\underline{0.159 (0.145, 0.174)} &\underline{0.115 (0.101, 0.129)}\\
  SmartPath-R1  &\textbf{0.259 (0.245, 0.274)} &\textbf{0.198 (0.184, 0.213)} &\textbf{0.133 (0.119, 0.147)}\\
\bottomrule
\end{tabular}
\end{table}

\begin{table}[htbp]
\centering
\caption{\textbf{External Validation of ROI-Level Detection Performance on GlaS Dataset.} 
Best performing model is \textbf{bolded} and second-best is \underline{underlined}. The 95\% CI is included in parentheses.}
\label{tab:ROI_det:GlaS}
\begin{tabular}{lccc}
\toprule
\textbf{Method} & \textbf{AP30} & \textbf{AP50} & \textbf{AP70} \\
  \midrule
  Qwen2.VL &0.119 (0.081, 0.157) & 0.068 (0.033, 0.102) &0.037 (0.006, 0.068)\\
  LLaVA-Med &0.002 (0.000, 0.005) &0.000 (0.000, 0.000) &0.000 (0.000, 0.000)\\
  Quilt-LLaVA  &0.090 (0.065, 0.114) &0.034 (0.015, 0.052) &0.015 (0.000, 0.030)\\ 
  PathoR1  &0.069 (0.046, 0.091) &0.028 (0.011, 0.046) &0.013 (0.000, 0.027) \\
  BiomedParse &0.057 (0.039, 0.074) &0.035 (0.024, 0.045) &0.020 (0.012, 0.028) \\
 SmartPath-SFT  &\underline{0.427 (0.367, 0.487)} &\underline{0.373 (0.310, 0.436)} &\underline{0.292 (0.230, 0.354)}\\
  SmartPath-R1  &\textbf{0.597 (0.537, 0.657)} &\textbf{0.533 (0.468, 0.598)} &\textbf{0.423 (0.353, 0.492)}\\
\bottomrule
\end{tabular}
\end{table}

\begin{table}[htbp]
\centering
\caption{\textbf{External Validation of ROI-Level Detection Performance on NuCLS Dataset.} 
Best performing model is \textbf{bolded} and second-best is \underline{underlined}. The 95\% CI is included in parentheses.}
\label{tab:ROI_det:NuCLS}
\begin{tabular}{lccc}
\toprule
\textbf{Method} & \textbf{AP30} & \textbf{AP50} & \textbf{AP70} \\
  \midrule
  Qwen2.VL &0.010 (0.006, 0.014) &0.005 (0.001, 0.008) & 0.001 (0.000, 0.003)\\
  LLaVA-Med &0.000 (0.000, 0.001) &0.000 (0.000, 0.000) &0.000 (0.000, 0.000)\\
  Quilt-LLaVA  & 0.000 (0.000, 0.000) &0.000 (0.000, 0.000) &0.000 (0.000, 0.000)\\  
  PathoR1  &  0.001 (0.000, 0.002) & 0.000 (0.000, 0.000) &  0.000 (0.000, 0.000)\\
  BiomedParse &0.114 (0.104, 0.124) &0.079 (0.071, 0.088) &0.047 (0.040, 0.054) \\
 SmartPath-SFT  &\underline{0.191 (0.173, 0.209)} &\underline{0.164 (0.147, 0.181)} &\underline{0.105 (0.091, 0.119)}\\
  SmartPath-R1  &\textbf{0.322 (0.303, 0.341)} &\textbf{0.262 (0.244, 0.280)} &\textbf{0.164 (0.148, 0.180)}\\
\bottomrule
\end{tabular}
\end{table}

\begin{table}[htbp]
\centering
\caption{\textbf{Average ROI-Level Segmentation Performance on 14 Dataset.} 
Best performing model is \textbf{bolded} and second-best is \underline{underlined}. The 95\% CI is included in parentheses.}
\label{tab:ROI_seg:avg}
\begin{tabular}{lc}
\toprule
\textbf{Method} & \textbf{Dice Score} \\
  \midrule
  Qwen2.VL & 0.090 (0.027, 0.154)\\
  LLaVA-Med & 0.117 (0.038, 0.196) \\
  Quilt-LLaVA  &0.295 (0.153, 0.436)  \\  
  PathoR1 &0.261 (0.147, 0.374)  \\
  BiomedParse &0.305 (0.154, 0.455)  \\
 SmartPath-SFT  &\underline{0.359 (0.255, 0.463)}  \\
  SmartPath-R1  &\textbf{0.500 (0.357, 0.642)} \\
\bottomrule
\end{tabular}
\end{table}

\begin{figure*}
    \centering
    \includegraphics[width=1\linewidth]{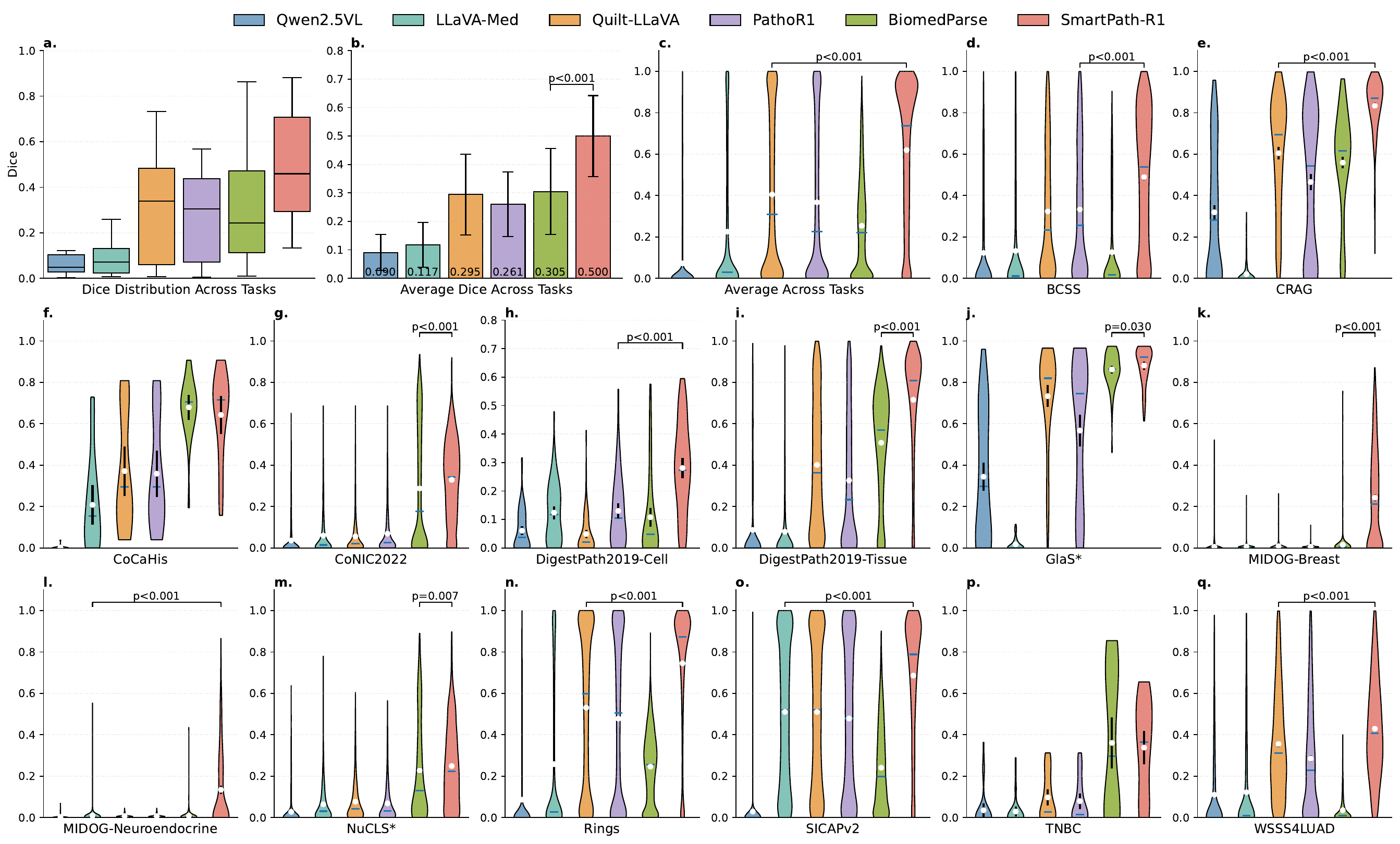}
\caption{\textbf{Performance of MLLMs on ROI-Level Segmentation Tasks.}
\textbf{a-c.} Average performance of MLLMs based on average Dice Score across 14 ROI-level segmentation tasks.
\textbf{d-q.} Model performance on specific tasks. * represents external validation datasets. The violin plots visualize the performance distribution of different methods, where each violin's width represents the data density. The blue horizontal line within each violin indicates the median value, while the white circle marks the mean point. Black vertical lines show the 95\% CI. P-values are computed using a Wilcoxon signed-rank two-sided test. 
}
\label{fig:ROI_seg_bar}
\end{figure*}

\begin{table}[htbp]
\centering
\caption{\textbf{ROI-Level Segmentation Performance on BCSS Dataset.} 
Best performing model is \textbf{bolded} and second-best is \underline{underlined}. The 95\% CI is included in parentheses.}
\label{tab:ROI_seg:BCSS}
\begin{tabular}{lcccc}
\toprule
\textbf{Method} & \textbf{BCSS} & \textbf{CRAG} & \textbf{CoCaHis} & \textbf{CoNIC2022} \\
  \midrule
  Qwen2.VL &0.123 (0.112, 0.133)&0.319 (0.285, 0.353) &0.002 (0.000, 0.005) &0.036 (0.033, 0.039) \\
  LLaVA-Med &0.132 (0.121, 0.143) &0.020 (0.015, 0.025) &0.208 (0.112, 0.304) &0.060 (0.057, 0.063) \\
  Quilt-LLaVA  &0.323 (0.310, 0.337) &\underline{0.604 (0.574, 0.635)} & 0.371 (0.251, 0.491) &0.056 (0.053, 0.058)\\  
  PathoR1 &0.333 (0.319, 0.347) &0.465 (0.427, 0.503) &0.359 (0.246, 0.471) &0.069 (0.066, 0.072)\\
  BiomedParse &0.126 (0.117, 0.136) &0.559 (0.531, 0.587) &\textbf{0.679 (0.619, 0.740)} &\underline{0.287 (0.278, 0.297)} \\
 SmartPath-SFT  &\underline{0.441 (0.425, 0.456)}  &0.474 (0.430, 0.517) &0.566 (0.448, 0.684)) &0.185 (0.178, 0.192)\\
  SmartPath-R1  &\textbf{0.489 (0.475, 0.504)} &\textbf{0.833 (0.818, 0.848)} &\underline{0.643 (0.552, 0.734)} &\textbf{0.329 (0.323, 0.335)}\\
    \midrule
& \textbf{DigestPath2019-Cell} & \textbf{DigestPath2019-Tissue} & \textbf{MIDOG-Breast} & \textbf{MIDOG-Neuroendocrine} \\
  \midrule
  Qwen2.VL &0.060 (0.045, 0.076) &0.087 (0.079, 0.095) &0.005 (0.004, 0.007)&0.003 (0.002, 0.004) \\
  LLaVA-Med &0.124 (0.101, 0.147) &0.078 (0.071, 0.085)&0.007 (0.006, 0.007) &0.009 (0.006, 0.012)\\
  Quilt-LLaVA  &0.049 (0.033, 0.064) &0.401 (0.390, 0.413) &0.008 (0.006, 0.010)  &0.007 (0.007, 0.008) \\   
    PathoR1 &0.130 (0.103, 0.157) &0.326 (0.314, 0.338) &0.006 (0.005, 0.006) &0.007 (0.006, 0.008)\\
  BiomedParse &0.108 (0.075, 0.141)  &0.508 (0.498, 0.518)& 0.015 (0.011, 0.018) &0.009 (0.006, 0.011) \\
 SmartPath-SFT  &\underline{0.130 (0.095, 0.165)}  &\underline{0.635 (0.624, 0.647)} &\underline{0.215 (0.197, 0.234)} &\underline{0.104 (0.083, 0.125)}\\
  SmartPath-R1  &\textbf{0.281 (0.245, 0.316)}  &\textbf{0.716 (0.706, 0.725)}&\textbf{0.243 (0.226, 0.260)} &\textbf{0.134 (0.114, 0.154)}\\
    \midrule
  &\textbf{Rings} & \textbf{SICAPv2}  & \textbf{TNBC}  & \textbf{WSSS4LUAD} \\
  \midrule
  Qwen2.VL &0.086 (0.083, 0.090)  &0.029 (0.025, 0.032)  & 0.035 (0.003, 0.067) &0.110 (0.099, 0.122)\\
  LLaVA-Med &0.258 (0.252, 0.264)  &0.510 (0.501, 0.520) & 0.029 (0.004, 0.053) &0.122 (0.110, 0.134)\\
  Quilt-LLaVA  &\underline{0.531 (0.525, 0.537)}  &\underline{0.510 (0.500, 0.520)} &0.097 (0.058, 0.137) &0.356 (0.341, 0.371)\\   
    PathoR1 &0.478 (0.472, 0.484) &0.478 (0.468, 0.489) &0.079 (0.042, 0.116) &0.285 (0.269, 0.300)\\
  BiomedParse &0.246 (0.243, 0.249) & 0.241 (0.235, 0.246) &\textbf{0.361 (0.237, 0.484)} & 0.036 (0.032, 0.039)\\
 SmartPath-SFT  &0.526 (0.519, 0.533) &0.416 (0.405, 0.427) &0.244 (0.164, 0.323)&\underline{0.356 (0.341, 0.372)} \\
  SmartPath-R1  &\textbf{0.745 (0.740, 0.750)}  &\textbf{0.686 (0.678, 0.694)}&\underline{0.338 (0.257, 0.419)} &\textbf{0.429 (0.415, 0.443)} \\
\bottomrule
\end{tabular}
\end{table}

\begin{table}[htbp]
\centering
\caption{\textbf{External Validation of ROI-Level Segmentation Performance on GlaS and NuCLS Dataset.} 
Best performing model is \textbf{bolded} and second-best is \underline{underlined}. The 95\% CI is included in parentheses.}
\label{tab:ROI_seg:GlaS}
\begin{tabular}{lcc}
\toprule
\textbf{Method} & \textbf{GlaS} & \textbf{NuCLS} \\
  \midrule
  Qwen2.VL & 0.344 (0.275, 0.413) &0.027 (0.022, 0.033)\\
  LLaVA-Med &0.017 (0.011, 0.024) &0.064 (0.058, 0.070) \\
  Quilt-LLaVA  &0.733 (0.680, 0.786) &0.077 (0.070, 0.083)  \\  
  PathoR1 &0.567 (0.491, 0.644) &0.069 (0.062, 0.075) \\
  BiomedParse &\underline{0.862 (0.843, 0.881)}  &\underline{0.227 (0.212, 0.243)} \\
 SmartPath-SFT  &0.548 (0.457, 0.639) &0.185 (0.171, 0.199) \\
  SmartPath-R1  &\textbf{0.881 (0.861, 0.901)} &\textbf{0.249 (0.236, 0.261)}\\
\bottomrule
\end{tabular}
\end{table}

\begin{table}[htbp]
\centering
\caption{\textbf{Average model performance across 10 pathological VQA datasets.} 
Best performing model for each dataset is \textbf{bolded} and second-best performing model is \underline{underlined}. 95\% CIs are shown in parentheses.}
\label{tab:ROI:avg_vqa}
\begin{tabular}{lc}
\toprule
 \textbf{Method}  &\textbf{ACC/BLEU/BERT-Score} \\
\midrule
  Qwen2.VL & 0.440 (0.305, 0.575) \\
  LLaVA-Med & 0.296 (0.138, 0.455) \\
 Quilt-LLaVA & 0.350 (0.210, 0.490) \\
   PathoR1 & 0.578 (0.415, 0.741) \\
  SmartPath-SFT & \underline{0.642 (0.514, 0.771)} \\
   SmartPath-R1 & \textbf{0.700 (0.574, 0.825)} \\
\bottomrule
\end{tabular}
\end{table}

\begin{table}[htbp]
\centering
\caption{\textbf{Model Performance across Multiple ROI-Level VQA Datasets.} 
Best performing model for each dataset is \textbf{bolded} and second-best performing model is \underline{underlined}. 95\% CIs are shown in parentheses.}
\label{tab:ROI:vqa}
\begin{tabular}{lcc}
\toprule
 \textbf{Method} & \textbf{Atlas-Test\_tiny} &\textbf{Atlas-Val} \\
\midrule
  Qwen2.VL  & 0.457 (0.388, 0.525)& 0.525 (0.413, 0.637)  \\
  LLaVA-Med  & 0.279 (0.217, 0.340)&0.287 (0.186, 0.389) \\
 Quilt-LLaVA  &0.462 (0.393, 0.530)&0.388 (0.278, 0.497) \\
 PathoR1  &\textbf{0.793 (0.738, 0.849)} & \textbf{0.812 (0.725, 0.900)}\\
  SmartPath-SFT  & 0.712 (0.649, 0.774) &0.763 (0.667, 0.858)\\
   SmartPath-R1  & \underline{0.764 (0.706, 0.823)} & \underline{0.800 (0.710, 0.890)} \\
\midrule
  & \textbf{EduContent-Test\_tiny} &\textbf{EduContent-Val} \\
\midrule
  Qwen2.VL  & 0.537 (0.476, 0.599)& 0.479 (0.397, 0.561) \\
  LLaVA-Med  & 0.220 (0.168, 0.271) & 0.171 (0.109, 0.233)\\
 Quilt-LLaVA  &0.396 (0.336, 0.457) &0.336 (0.258, 0.413) \\
  PathoR1  & \underline{0.710 (0.654, 0.766)} &0.603 (0.522, 0.683) \\
  SmartPath-SFT  & 0.694 (0.637, 0.751) &\underline{0.616 (0.537, 0.696)}\\
   SmartPath-R1  & \textbf{0.729 (0.675, 0.784)} & \textbf{0.651 (0.572, 0.729)} \\
\midrule
  & \textbf{PathCLS-Test\_tiny} &\textbf{PathCLS-Val} \\
\midrule
  Qwen2.VL & 0.309 (0.238, 0.380) & 0.233 (0.144, 0.322)\\
  LLaVA-Med & 0.188 (0.128, 0.248) & 0.178 (0.097, 0.258)\\
 Quilt-LLaVA & 0.145 (0.091, 0.200)&  0.133 (0.062, 0.205) \\
  PathoR1  & 0.376 (0.301, 0.450) &0.378 (0.276, 0.480)\\
  SmartPath-SFT & \underline{0.588 (0.512, 0.664)} & \underline{0.611 (0.508, 0.714)}\\
   SmartPath-R1 & \textbf{0.721 (0.652, 0.790)}& \textbf{0.722 (0.628, 0.817)}  \\
\midrule
  & \textbf{PubMed-Test\_tiny} &\textbf{PubMed-Val} \\
\midrule
  Qwen2.VL &0.552 (0.493, 0.610) & 0.472 (0.408, 0.537)\\
  LLaVA-Med & 0.210 (0.162, 0.258) &0.232 (0.177, 0.286) \\
 Quilt-LLaVA & 0.391 (0.334, 0.449) &0.356 (0.294, 0.418) \\
  PathoR1  &\underline{0.669 (0.614, 0.724)} & \underline{0.618 (0.555, 0.681)}\\
  SmartPath-SFT  & 0.665 (0.610, 0.721) & 0.541 (0.476, 0.605) \\
   SmartPath-R1  & \textbf{0.737 (0.685, 0.788)} & \textbf{0.652 (0.591, 0.714)}\\
   \midrule
     & \textbf{PathVQA-Close} &\textbf{PathVQA-Open, BLEU} \\
\midrule
  Qwen2.VL &0.490 (0.473, 0.506) & 0.001 (0.001, 0.001)\\
  LLaVA-Med &  0.684 (0.668, 0.699) &0.002 (0.002, 0.003) \\
 Quilt-LLaVA & 0.447 (0.430, 0.464) &0.001 (0.001, 0.001) \\
  PathoR1  &0.605 (0.588, 0.622) &0.002 (0.002, 0.002)\\
  SmartPath-SFT  & \underline{0.854 (0.842, 0.866)} & \underline{0.158 (0.147, 0.170)} \\
   SmartPath-R1  & \textbf{0.864 (0.852, 0.875)} & \textbf{0.182 (0.169, 0.194)}\\
      \midrule
     & \textbf{PathVQA-Open, BERT-Score} &\\
\midrule
  Qwen2.VL &0.782 (0.782, 0.783) &\\
  LLaVA-Med &  0.808 (0.807, 0.808) & \\
 Quilt-LLaVA & 0.795 (0.794, 0.796) & \\
  PathoR1  &0.794 (0.794, 0.795) &\\
  SmartPath-SFT  & \underline{0.865 (0.862, 0.868)} & \\
   SmartPath-R1  & \textbf{0.875 (0.872, 0.877)} & \\
\bottomrule
\end{tabular}
\end{table}

\begin{table}[htbp]
\centering
\caption{\textbf{Classes of Different Datasets for WSI-Level Classification.} * represents external validation datasets.
}\label{tab:WSI:classes_cls}%
\begin{tabular}{lc}
\toprule
 \textbf{Dataset} & \textbf{Class}  \\
 \midrule
 TCGA-BRCA &Invasive ductal carcinoma; Invasive lobular carcinoma; Invasive mucinous carcinoma; \\ &Invasive papillary carcinoma; Invasive micropapillary carcinoma\\
  TCGA-COAD &Adenocarcinoma, NOS; Mucinous adenocarcinoma; Tubular adenocarcinoma; \\ &Serrated adenocarcinoma; Signet ring cell carcinoma\\
 TCGA-HNSC &Invasive well-differentiated squamous carcinoma; Invasive moderately differentiated squamous carcinoma; 
 \\& Invasive poorly differentiated squamous carcinoma; Non-invasive squamous carcinoma\\
 TCGA-LGG &Anaplastic astrocytoma; Oligodendroglioma; Diffuse astrocytoma;  Glioblastoma multiforme\\
 TCGA-LUAD &Lepidic; Acinar; Papillary; Micropapillary; Solid\\
 TCGA-LUSC &Keratinizing squamous cell carcinoma; Non-keratinizing squamous cell carcinoma;\\ &Basaloid squamous cell carcinoma; Verrucous squamous cell carcinoma\\
BRACS* 
&Normal; Pathological benign; Usual ductal hyperplasis; Flat epithelial atypia;\\ 
&Atypical ductal hyperplasia; Ductal carcinoma in situ; Invasive carcinoma \\
CAMELYON* &Normal; Tumor \\
CPTAC-LSCC* & Lung adenocarcinoma; Lung squamous cell carcinoma \\
\bottomrule
\end{tabular}
\end{table}

\begin{table}[htbp]
\centering
\caption{\textbf{Average WSI-level Classification Performance on 9 datasets.} 
Best performing model is \textbf{bolded} and second-best is \underline{underlined}. The 95\% CI is included in parentheses.}
\label{tab:slide_cls:avg}
\begin{tabular}{lc}
\toprule
\textbf{Method} & \textbf{ACC} \\
  \midrule
  Qwen2.VL &0.353 (0.225, 0.482) \\
  LLaVA-Med & 0.306 (0.176, 0.436) \\
  Quilt-LLaVA  & 0.274 (0.179, 0.369) \\  
  PathoR1 &0.551 (0.410, 0.692) \\
 SmartPath-SFT  &\underline{0.698 (0.538, 0.858)}  \\
  SmartPath-R1  &\textbf{0.706 (0.555, 0.856)} \\
\bottomrule
\end{tabular}
\end{table}

\begin{table}[htbp]
\centering
\caption{\textbf{WSI-level Classification Performance on TCGA-BRCA, TCGA-COAD, TCGA-HNSC, TCGA-LGG, TCGA-LUAD and TCGA-LUSC datasets.} 
Best performing model is \textbf{bolded} and second-best is \underline{underlined}. The 95\% CI is included in parentheses.}
\label{tab:WSI:tcga}
\begin{tabular}{lcccc}
\toprule
\textbf{Method} &Dataset &\textbf{\# Class} & \textbf{ACC} & \textbf{F1}  \\
\midrule
Qwen2.VL &TCGA-BRCA &5  & 0.359 (0.259, 0.459)&0.340 (0.268,0.499) \\
LLaVA-Med &TCGA-BRCA &5   & 0.217 (0.132, 0.303)&0.167 (0.156,0.318) \\
Quilt-LLaVA &TCGA-BRCA &5    & 0.152 (0.077, 0.227) & 0.226 (0.166,0.345)\\
PathoR1 &TCGA-BRCA &5    & 0.533 (0.429, 0.637)&0.137 (0.084,0.209) \\
SmartPath-SFT &TCGA-BRCA &5   & \underline{0.674 (0.576, 0.772)}& \underline{0.626 (0.478,0.751)}\\
SmartPath-R1 &TCGA-BRCA &5   & \textbf{0.750 (0.660, 0.840)} &\textbf{0.781 (0.578,0.825)}\\
\midrule
Qwen2.VL &TCGA-COAD &5  & 0.568 (0.400, 0.735)&0.342 (0.365,0.613) \\
LLaVA-Med &TCGA-COAD &5   & 0.432 (0.265, 0.600)&0.268 (0.128,0.354) \\
Quilt-LLaVA &TCGA-COAD &5    & 0.351 (0.190, 0.513) &0.375 (0.157,0.424) \\
PathoR1 &TCGA-COAD &5    & 0.757 (0.612, 0.902)& 0.402 (0.165,0.439)\\
SmartPath-SFT &TCGA-COAD &5   & \underline{0.838 (0.713, 0.962)}& \underline{0.808 (0.550,0.948)}\\
SmartPath-R1 &TCGA-COAD &5   & \textbf{0.838 (0.713, 0.962)} &\textbf{0.977 (0.577,0.940)}\\
\midrule
Qwen2.VL  &TCGA-HNSC &4  & 0.480 (0.270, 0.690)&0.431 (0.235,0.569) \\
LLaVA-Med &TCGA-HNSC  &4   & 0.280 (0.091, 0.469)& 0.156 (0.095,0.393)\\
Quilt-LLaVA &TCGA-HNSC  &4    & 0.440 (0.231, 0.649) & 0.237 (0.124,0.486)\\
PathoR1 &TCGA-HNSC  &4    & 0.520 (0.310, 0.730)& 0.261 (0.068,0.339)\\
SmartPath-SFT &TCGA-HNSC  &4   & \underline{0.800 (0.631, 0.969)}& \textbf{0.801 (0.516,0.937)}\\
SmartPath-R1 &TCGA-HNSC  &4   & \textbf{0.800 (0.631, 0.969)} &\underline{0.681 (0.652,0.944)}\\
\midrule
Qwen2.VL &TCGA-LGG &4  & 0.216 (0.077, 0.355)&0.066 (0.042,0.226) \\
LLaVA-Med &TCGA-LGG &4   & 0.189 (0.057, 0.322)&0.093 (0.054,0.230) \\
Quilt-LLaVA &TCGA-LGG &4    &0.216 (0.077, 0.355) &0.215 (0.097,0.319) \\
PathoR1 &TCGA-LGG &4    & 0.595 (0.429, 0.761)&0.169 (0.055,0.255) \\
SmartPath-SFT &TCGA-LGG  &4   & \underline{0.811 (0.678, 0.943)}& \textbf{0.722 (0.470,0.883)}\\
SmartPath-R1 &TCGA-LGG &4   & \textbf{0.892 (0.787, 0.997)} &\underline{0.715 (0.544,0.981)}\\
\midrule
Qwen2.VL &TCGA-LUAD &5  & 0.520 (0.377, 0.663)&0.473 (0.307,0.594) \\
LLaVA-Med &TCGA-LUAD &5   & 0.300 (0.168, 0.432)& 0.228 (0.081,0.303)\\
Quilt-LLaVA &TCGA-LUAD &5    & 0.140 (0.040, 0.240) &0.193 (0.102,0.297) \\
PathoR1 &TCGA-LUAD &5    & 0.760 (0.637, 0.883)&0.155 (0.074,0.267) \\
SmartPath-SFT &TCGA-LUAD &5   & \underline{0.840 (0.735, 0.945)}& \underline{0.689 (0.545,0.866)}\\
SmartPath-R1 &TCGA-LUAD &5   & \textbf{0.860 (0.760, 0.960)} &\textbf{0.715 (0.573,0.911)}\\
\midrule
Qwen2.VL &TCGA-LUSC &4  &  0.160 (0.055, 0.265)&0.094 (0.050,0.202) \\
LLaVA-Med &TCGA-LUSC &4   & 0.120 (0.027, 0.213)&0.265 (0.198,0.400) \\
Quilt-LLaVA &TCGA-LUSC &4    & 0.280 (0.151, 0.409)&0.304 (0.163,0.413) \\
PathoR1 &TCGA-LUSC &4    & 0.620 (0.481, 0.759)& 0.180 (0.128,0.368)\\
SmartPath-SFT &TCGA-LUSC &4   & \textbf{0.920 (0.842, 0.998)}& \textbf{0.901 (0.747,0.987)}\\
SmartPath-R1 &TCGA-LUSC 
&4   & \underline{0.760 (0.637, 0.883)} &\underline{0.681 (0.580,0.822)}\\
\bottomrule
\end{tabular}
\end{table}

\begin{table}[htbp]
\centering
\caption{\textbf{External Validation of WSI-Level Classification Performance on BRACS Dataset.} 
Best performing model is \textbf{bolded} and second-best is \underline{underlined}. The 95\% CI is included in parentheses.}
\label{tab:WSI:BRACS}
\begin{tabular}{lccc}
\toprule
\textbf{Method} &\textbf{\# Class} & \textbf{ACC} & \textbf{F1}  \\
\midrule
Qwen2.VL &7  &  0.110 (0.084, 0.136) &0.092 (0.072,0.118) \\
LLaVA-Med &7   & 0.119 (0.092, 0.147) & \textbf{0.132 (0.104,0.158)}\\
Quilt-LLaVA &7    & 0.110 (0.084, 0.136) &0.111 (0.101,0.150) \\
PathoR1 &7    & 0.138 (0.109, 0.167)& 0.122 (0.100,0.157)\\
SmartPath-SFT &7   & \underline{0.250 (0.213, 0.286)}& 0.077 (0.059,0.094)\\
SmartPath-R1 &7   & \textbf{0.272 (0.234, 0.309)} &\underline{0.124 (0.098,0.139)}\\
\bottomrule
\end{tabular}
\end{table}

\begin{table}[htbp]
\centering
\caption{\textbf{External Validation of WSI-Level Classification Performance on CAMELYON Dataset.} 
Best performing model is \textbf{bolded} and second-best is \underline{underlined}. The 95\% CI is included in parentheses.}
\label{tab:WSI:CAMELYON}
\begin{tabular}{lccc}
\toprule
\textbf{Method} &\textbf{\# Class} & \textbf{ACC} & \textbf{F1}  \\
\midrule
Qwen2.VL &2  &  0.477 (0.444, 0.509) &\textbf{0.511 (0.455,0.556)} \\
LLaVA-Med &2   & \underline{0.589 (0.557, 0.621)} & 0.447 (0.406,0.510)\\
Quilt-LLaVA &2    &  0.381 (0.349, 0.413) & 0.505 (0.442,0.535)\\
PathoR1 &2    & 0.541 (0.509, 0.574) & \underline{0.506 (0.445,0.540)}\\
SmartPath-SFT &2   & \textbf{0.620 (0.588, 0.652)}& 0.390 (0.372,0.404)\\
SmartPath-R1 &2   & \textbf{0.620 (0.588, 0.652)} &0.388 (0.373,0.404)\\
\bottomrule
\end{tabular}
\end{table}

\begin{table}[htbp]
\centering
\caption{\textbf{External Validation of WSI-Level Classification Performance on CPTAC-NSCLC Dataset.} 
Best performing model is \textbf{bolded} and second-best is \underline{underlined}. The 95\% CI is included in parentheses.}
\label{tab:WSI:CPTAC-NSCLC}
\begin{tabular}{lccc}
\toprule
\textbf{Method} &\textbf{\# Class} & \textbf{ACC} & \textbf{F1}  \\
\midrule
Qwen2.VL &2  &  0.290 (0.271, 0.309) &0.490 (0.471,0.511 \\
LLaVA-Med &2   &0.507 (0.486, 0.527) &0.505 (0.481,0.523) \\
Quilt-LLaVA &2    & 0.392 (0.372, 0.413) & 0.500 (0.473,0.514)\\
PathoR1 &2    & 0.495 (0.474, 0.516) & 0.475 (0.472,0.514)\\
SmartPath-SFT &2   & \underline{0.529 (0.508, 0.550)}& \underline{0.508 (0.505,0.547)}\\
SmartPath-R1 &2   & \textbf{0.558 (0.537, 0.579)} &\textbf{0.556 (0.538,0.578)}\\
\bottomrule
\end{tabular}
\end{table}

\begin{table}[htbp]
\centering
\caption{\textbf{Average WSI-Level VQA Performance on 17 Datasets.} 
Best performing model is \textbf{bolded} and second-best is \underline{underlined}. The 95\% CI is included in parentheses.}
\label{tab:slide_vqa:avg}
\begin{tabular}{lc}
\toprule
\textbf{Method} & \textbf{ACC} \\
  \midrule
  Qwen2.VL & 0.360 (0.302, 0.417)\\
  LLaVA-Med & 0.309 (0.246, 0.372) \\
  Quilt-LLaVA  &0.350 (0.301, 0.398)  \\  
  PathoR1 &0.505 (0.427, 0.582)  \\
 SmartPath-SFT  &\underline{0.684 (0.592, 0.776)}  \\
  SmartPath-R1  &\textbf{0.695 (0.604, 0.786)} \\
\bottomrule
\end{tabular}
\end{table}

\begin{table}[htbp]
\centering
\caption{\textbf{WSI-Level VQA Performance on Microscopy Tasks of TCGA Dataset.} 
Best performing model is \textbf{bolded} and second-best is \underline{underlined}. The 95\% CI is included in parentheses.}
\label{tab:slide_vqa:microscopy}
\begin{tabular}{lcc}
\toprule
\textbf{Method} & \textbf{Cytomorphological Characteristics} & \textbf{Histopathological Changes} \\
  \midrule
  Qwen2.VL &0.475 (0.313, 0.637)&0.322 (0.236, 0.408) \\
  LLaVA-Med &0.150 (0.034, 0.266) & 0.271 (0.190, 0.353) \\
  Quilt-LLaVA  &0.425 (0.265, 0.585) &0.407 (0.317, 0.497)\\   
  PathoR1 &0.700 (0.552, 0.848) & 0.568 (0.477, 0.658)\\
 SmartPath-SFT  &\underline{0.750 (0.610, 0.890)} &\underline{0.805 (0.733, 0.878)} \\
  SmartPath-R1  &\textbf{0.825 (0.702, 0.948)} & \textbf{0.822 (0.752, 0.892)}\\
    \midrule
  \textbf{Method}  & \textbf{Tissue Architecture and Arrangement} & \textbf{Tumor Characteristics}  \\
  \midrule
  Qwen2.VL &0.447 (0.366, 0.527) & 0.301 (0.200, 0.402)\\
  LLaVA-Med &0.393 (0.314, 0.472) &0.217 (0.126, 0.307)\\
  Quilt-LLaVA  &0.453 (0.373, 0.534) &0.446 (0.337, 0.555) \\  
    PathoR1 &0.660 (0.583, 0.737) & 0.614 (0.508, 0.721)\\
 SmartPath-SFT  &\textbf{0.807 (0.743, 0.871)} &\underline{0.747 (0.651, 0.842)}\\
  SmartPath-R1  &\underline{0.800 (0.735, 0.865)} &\textbf{0.771 (0.679, 0.863)} \\
\bottomrule
\end{tabular}
\end{table}

\begin{table}[htbp]
\centering
\caption{\textbf{WSI-Level VQA Performance on Diagnosis Tasks of TCGA Dataset.} 
Best performing model is \textbf{bolded} and second-best is \underline{underlined}. The 95\% CI is included in parentheses.}
\label{tab:slide_vqa:diagnosis}
\begin{tabular}{lccc}
\toprule
\textbf{Method} & \textbf{Differential Diagnosis} & \textbf{Disease Classification}  & \textbf{Disease Detection}\\
  \midrule
  Qwen2.VL &0.269 (0.145, 0.394)&0.393 (0.342, 0.444)&0.409 (0.186, 0.632) \\
  LLaVA-Med &0.212 (0.097, 0.326) &0.259 (0.214, 0.305)&0.545 (0.319, 0.771)\\
  Quilt-LLaVA  &0.250 (0.128, 0.372) &0.253 (0.208, 0.299)&0.409 (0.186, 0.632)\\   
    PathoR1 &\underline{0.635 (0.499, 0.770)}& 0.632 (0.582, 0.682) &0.500 (0.273, 0.727)\\
 SmartPath-SFT &\textbf{0.731 (0.606, 0.855)} &\underline{0.805 (0.764, 0.846)} &\textbf{0.727 (0.525, 0.929)} \\
  SmartPath-R1  &\textbf{0.731 (0.606, 0.855)} &\textbf{0.808 (0.767, 0.849)}&\underline{0.591 (0.368, 0.814)}\\
    \midrule
    \textbf{Method} & \textbf{Grading} & \textbf{Staging}  \\
    \midrule
  Qwen2.VL &0.383 (0.326, 0.439) &0.308 (0.254, 0.363) & \\
  LLaVA-Med &0.228 (0.179, 0.276) &0.197 (0.150, 0.244) &\\
  Quilt-LLaVA  &0.331 (0.277, 0.386) &0.237 (0.186, 0.287) &\\  
    PathoR1 &0.403 (0.347, 0.460) &  0.380 (0.323, 0.437) &\\
 SmartPath-SFT &\underline{0.617 (0.561, 0.674)} &\textbf{0.670 (0.615, 0.726)} & \\
  SmartPath-R1  &\textbf{0.666 (0.611, 0.720)} &\underline{0.652 (0.596, 0.709)} &\\
\bottomrule
\end{tabular}
\end{table}

\begin{table}[htbp]
\centering
\caption{\textbf{WSI-Level VQA Performance on Clinical Tasks of TCGA Dataset.} 
Best performing model is \textbf{bolded} and second-best is \underline{underlined}. The 95\% CI is included in parentheses.}
\label{tab:slide_vqa:clinical}
\begin{tabular}{lcc}
\toprule
\textbf{Method} & \textbf{Biomarker Analysis} & \textbf{Prognostic Assessment} \\
  \midrule
  Qwen2.VL &0.167 (0.000, 0.414)&0.323 (0.148, 0.497) \\
  LLaVA-Med &0.167 (0.000, 0.414) & 0.290 (0.121, 0.460) \\
  Quilt-LLaVA  &0.250 (0.000, 0.537) &0.484 (0.298, 0.670)\\  
    PathoR1 & 0.417 (0.089, 0.744) &0.484 (0.298, 0.670) \\
 SmartPath-SFT  &\underline{0.500 (0.168, 0.832)} &\textbf{0.806 (0.659, 0.954)} \\
  SmartPath-R1  &\textbf{0.667 (0.354, 0.980)} & \underline{0.742 (0.579, 0.905)}\\
    \midrule
  \textbf{Method}  & \textbf{Risk Factors} & \textbf{Treatment Guidances}  \\
  \midrule
  Qwen2.VL &0.364 (0.025, 0.703) & 0.638 (0.496, 0.781)\\
  LLaVA-Med &0.545 (0.195, 0.896) &0.489 (0.341, 0.638)\\
  Quilt-LLaVA  &0.273 (0.000, 0.587) &0.468 (0.320, 0.616) \\   
    PathoR1 &\underline{0.727 (0.455, 1.000)} &0.553 (0.406, 0.701) \\
 SmartPath-SFT  & \textbf{0.909 (0.818, 1.000)}&\underline{0.830 (0.718, 0.941)}\\
  SmartPath-R1  & \textbf{0.909 (0.818, 1.000)} &\textbf{0.894 (0.802, 0.985)} \\
\bottomrule
\end{tabular}
\end{table}

\begin{table}[htbp]
\centering
\caption{\textbf{External Validation of WSI-Level VQA Performance on BCNB Dataset.} 
Best performing model is \textbf{bolded} and second-best is \underline{underlined}. The 95\% CI is included in parentheses.}
\label{tab:slide_vqa:bcnb}
\begin{tabular}{lcccc}
\toprule
\textbf{Method}  & \textbf{Grading}  & \textbf{HER2 Expression} &\textbf{HER2 Type}&\textbf{Tumor Type}\\
  \midrule
  Qwen2.VL &0.327 (0.297, 0.357) &\underline{0.235 (0.210, 0.261)} &0.263 (0.236, 0.289) &0.490 (0.459, 0.520)\\
  LLaVA-Med  &0.352 (0.321, 0.383) &0.231 (0.205, 0.256) &0.313 (0.285, 0.341) & 0.313 (0.285, 0.341)\\
  Quilt-LLaVA   & 0.319 (0.289, 0.349) &\textbf{0.239 (0.213, 0.265)}&0.262 (0.235, 0.288)& 0.364 (0.335, 0.393) \\   
    PathoR1 &\textbf{0.471 (0.439, 0.503)}&0.194 (0.170, 0.218) &0.317 (0.289, 0.345) &0.324 (0.296, 0.352)\\
 SmartPath-SFT  &0.422 (0.390, 0.454) &0.177 (0.154, 0.200) &\underline{0.712 (0.684, 0.739)} & \textbf{0.610 (0.580, 0.639)}\\
  SmartPath-R1   &\underline{0.435 (0.403, 0.467)}&0.197 (0.173, 0.221) &\textbf{0.730 (0.703, 0.756)} & \underline{0.578 (0.548, 0.607)}\\
\bottomrule
\end{tabular}
\end{table}

\begin{table}[]
    \centering
    \caption{The Number of Samples at WSI-Level Used for Training and Evaluation.}
    \begin{tabular}{lcccc}
        \toprule
        \textbf{Dataset}  & \textbf{Anatomical Region} & \textbf{\# Training Samples} & \textbf{\# Evaluation Samples} & \textbf{\# Samples} \\
        \midrule
        TCGA-LUAD  & Lung  & 18,390     & 191  & 18,581 \\
        TCGA-GBM  & Brain  & 21,495    & 20   & 21,515 \\
        TCGA-LUSC  & Lung   & 16,826     & 177  & 17,003 \\
        TCGA-BLCA & Bladder   & 15,653     & 158  & 15,811 \\
        TCGA-BRCA & Breast   & 38,427     & 452  & 38,879 \\
        TCGA-COAD  & Colon  & 14,756     & 157  & 14,913 \\
        TCGA-READ  & Rectum & 5,413     & 76   & 5,489 \\
        TCGA-HNSC & Head and Neck   & 13,994     & 99   & 14,093 \\
        TCGA-SKCM  & Skin  & 4,154     & 34   & 4,188 \\
        TCGA-LGG  & Brain  & 30,789     & 137  & 30,926 \\
        BRCAS &Breast &- &545 &545 \\
        CAMELYON &Breast &- &898 &898 \\
        CPTAC-NSCLC &Lung &- &1,213 &1,213 \\
        BCNB-Grading & Breast & - & 926 & 926 \\
        BCNB-HER2 Expression & Breast & - & 1,058 & 1,058  \\
        BCNB-HER2 Type & Breast & - & 1,058 & 1,058 \\
        BCNB-Tumor Type & Breast & - & 1,058 & 1,058 \\
        \midrule
        Total & & 179,897 &8,257  &188,154 \\
        \bottomrule
    \end{tabular}
    \label{tab:data_details_WSI}
\end{table}

\begin{table}[htbp]
    \centering
    \caption{The Number of Samples at ROI-Level Used for Training and Evaluation.}
    \begin{tabular}{lcccc}
        \toprule
        \textbf{Dataset}  &\textbf{Anatomical Region} & \textbf{\# Training Samples} & \textbf{\# Evaluation Samples} & \textbf{\# Samples} \\
        \midrule
        PathCap &Unspecified&207,000 &-&207,000\\
      PathInstruct &Unspecified &180,000 &-&180,000\\
      Quilt-1M &Unspecified&1,017,712 &-&1,017,712\\
    CCRCC &Kidney   & 22,530    &5,635 &28,165\\
    Chaoyang &Colon &4,021 &2,139&6,160\\
    CRC-MSI &Colon  &19,557 &32,361&51,918\\
    ESCA &Esophagus&178,187&189,142&367,329\\
    Pancancer-TIL &Unspecified&209,221&56,275&265,496\\
    UniToPatho &Colon &6,270&2,399&8,669\\
    PathMMU-PubMed &Unspecified&2,787 &514 &3,301\\
    PathMMU-Atlas &Unspecified&799 &288 &1,087\\
    PathMMU-EduContent &Unspecified&1,683 &401 &2,084\\
        PathMMU-PathCLS &Unspecified &1,632 &257 &1,889\\
    PathVQA &Unspecified &19,654 & 6,719&26,373\\
    BCSS & Breast &7,322 &1,870&9,192\\
    CoCaHis &Liver &58 &24&82\\
    CoNIC2022  &Colon &15,387 &3,945&19,332\\
    CRAG  &Colon &1,429 &321&1,750\\
    DigestPath2019 Cell  &Colon &352 &85&437\\
    DigestPath2019 Tissue  &Colon &10,725 &2,666&13,391\\
    MIDOG Breast   &Breast &3,032 &762 &3,794\\
    MIDOG Neuroendocrine & Pancreas&1,448 &354 &1,802\\
    Rings &Prostate &29,451 & 14,288 &43,739\\
    SICAPv2 &Prostate &19,140 &4,784 &23,924\\
    TNBC &Breast &134 &33&167\\
    WSS4LUAD &Lung &4,590 &1,145 &5,735\\
    BreakHis  &Breast &- &1,582 &1,582\\
    PanCancer-TCGA &Unspecified &- &2,000 &2,000\\
    GlaS   &Colon&- &80&80\\
    NuCLS &Breast &- &905&905\\
        \midrule
        Total & & 1,964,121 & 330,974  &2,295,095 \\
        \bottomrule
    \end{tabular}
    \label{tab:data_details_ROI}
\end{table}

\begin{table}[htbp]
\caption{\textbf{The Public Datasets Used in This Study.} Please note that some datasets may need permission for access.}
    \begin{tabular}{ll}
        \toprule
         \pmb{Dataset}& \pmb{Link or Source}\\
         \midrule
    TCGA \cite{weinstein2013cancer}&\url{https://portal.gdc.cancer.gov/} \\
    CPTAC \cite{edwards2015cptac}&\url{https://proteomic.datacommons.cancer.gov/pdc/}\\
    BRCAS \cite{brancati2022bracs} &\url{https://www.bracs.icar.cnr.it/download/} \\
    CAMELYON16 \cite{bejnordi2017diagnostic}&\url{https://camelyon16.grand-challenge.org/Data/} \\
    CAMELYON17 \cite{bandi2018detection}&\url{https://camelyon17.grand-challenge.org/Data/}\\
    BCNB \cite{xu2021predicting}&\url{https://bcnb.grand-challenge.org/}\\
    SlideInstruction \cite{chen2025slidechat} &\url{https://huggingface.co/datasets/General-Medical-AI/SlideChat}\\
 PathCap \cite{pathasst} &\url{https://huggingface.co/datasets/jamessyx/PathCap}\\
  PathInstruct \cite{pathasst} &\url{https://huggingface.co/datasets/jamessyx/PathInstruct}\\
 Quilt-1M \cite{ikezogwo2023quilt} &\url{https://github.com/wisdomikezogwo/quilt1m}\\
    CCRCC \cite{brummer2023computational} &\url{https://zenodo.org/records/7898308}\\
    Chaoyang \cite{zhu2021hard}&\url{https://github.com/bupt-ai-cz/HSA-NRL}\\
    CRC-MSI \cite{kather2019deep}&\url{https://zenodo.org/records/3832231}\\
    ESCA\cite{tolkach2023artificial}&\url{https://zenodo.org/records/7548828}\\
    PanCancer-TIL \cite{abousamra2022deep}&\url{https://zenodo.org/records/6604094}\\
    UniToPatho \cite{barbano2021unitopatho}&\url{https://ieee-dataport.org/open-access/unitopatho}\\
    BreakHis  \cite{spanhol2015dataset}&\url{https://www.kaggle.com/datasets/ambarish/breakhis}\\
    PanCancer-TCGA \cite{komura2022universal}&\url{https://zenodo.org/records/5889558}\\
    PathMMU \cite{sun2024pathmmu} &\url{https://huggingface.co/datasets/jamessyx/PathMMU}\\
    PathVQA \cite{he2020pathvqa}&\url{https://github.com/UCSD-AI4H/PathVQA}\\
    BCSS \cite{amgad2019structured} &\url{https://bcsegmentation.grand-challenge.org/}\\
    CoCaHis \cite{sitnik2021dataset} &\url{https://cocahis.irb.hr/} \\
    CoNIC2022 \cite{graham2024conic} &\url{https://conic-challenge.grand-challenge.org/} \\
    CRAG \cite{graham2019mild}  &\url{https://warwick.ac.uk/fac/cross_fac/tia/data/mildnet/} \\
    DigestPath2019 Cell \cite{da2022digestpath}  &\url{https://digestpath2019.grand-challenge.org/} \\
    DigestPath2019 Tissue \cite{da2022digestpath}  &\url{https://digestpath2019.grand-challenge.org/} \\
    GlaS \cite{sirinukunwattana2017gland} &\url{https://warwick.ac.uk/fac/cross_fac/tia/data/glascontest/} \\
        MIDOG Breast \cite{aubreville2024domain} &\url{https://midog2022.grand-challenge.org/}\\
    MIDOG Neuroendocrine \cite{aubreville2024domain} &\url{https://midog2022.grand-challenge.org/}\\
    NuCLS \cite{amgad2022nucls} &\url{https://sites.google.com/view/nucls/home} \\
    Rings \cite{salvi2021hybrid}  &\url{https://data.mendeley.com/datasets/h8bdwrtnr5/1} \\
    SICAPv2 \cite{silva2020going}&\url{https://data.mendeley.com/datasets/9xxm58dvs3/1}\\
    TNBC \cite{naylor2018segmentation}  &\url{https://zenodo.org/records/1175282#.YMisCTZKgow}\\
    WSS4LUAD \cite{han2022wsss4luad} &\url{https://wsss4luad.grand-challenge.org/}\\
         \bottomrule
    \end{tabular}
    \label{tab:data_links}
\end{table}

\end{document}